\documentclass[preprint2]{aastex}

\newcommand{\mathbold}[1]{\mbox{\protect\boldmath $#1$}}

\makeatletter
\renewcommand\appendix{%
 \par 
 \setcounter{section}{\z@}%
 \setcounter{subsection}{\z@}%
 \setcounter{equation}{\z@}%
 \def\thesection{\Alph{section}}%
 \def\theequation{%
  \thesection\arabic{equation}%
 }%
 \@addtoreset{equation}{section}%
 \appendix@figtab@defs 
 \def\section{%
  \@startsection 
   {section}{1}{\z@}%
   {5ex\@plus.5ex}{1ex\@plus.2ex}{\normalsize\bfseries}%
   }%
}%
\makeatother

\def\lra{\leftrightarrow}
\def\be{\begin{equation}}
\def\ee{\end{equation}}
\def\bea{\begin{eqnarray}}
\def\eea{\end{eqnarray}}
\newcommand{\lsim}{\; ^< \!\!\!\! _\sim \;}
\newcommand{\gsim}{\; ^> \!\!\!\! _\sim \;}
\newcommand{\abs}[1]{\left |\right . #1 \left .\right |}

\def\re{{\rm e}}
\def\rp{{\rm p}}
\def\rn{{\rm n}}
\def\rN{{\rm N}}
\def\nue{\nu_{\rm e}}
\def\Ye{Y_{\rm e}}

\begin{document}

\title{\uppercase{Monte Carlo Study of Supernova 
Neutrino Spectra Formation}}

\author{Mathias Th.~Keil and Georg G.~Raffelt} 
\affil{Max-Planck-Institut f\"ur Physik 
(Werner-Heisenberg-Institut)\\
F\"ohringer Ring 6, 80805 M\"unchen, Germany}

\author{Hans-Thomas Janka} 
\affil{Max-Planck-Institut f\"ur Astrophysik\\
Karl-Scharzschild-Str.~1, 85741 Garching, Germany}

\begin{abstract}
The neutrino flux and spectra formation in a supernova core is studied
by using a Monte Carlo code. The dominant opacity contribution for
$\nu_\mu$ is elastic scattering on nucleons
$\nu_{\mu}\rN\to\rN\nu_{\mu}$, where $\nu_\mu$ always stands for
either $\nu_\mu$ or $\nu_\tau$.  In addition we switch on or off a
variety of processes which allow for the exchange of energy or the
creation and destruction of neutrino pairs, notably nucleon
bremsstrahlung $\rN\rN\to\rN\rN\nu_{\mu}\bar\nu_{\mu}$, the pair
annihilation processes $\re^+\re^-\to\nu_{\mu}\bar\nu_{\mu}$ and
$\nue\bar\nue\to\nu_{\mu}\bar\nu_{\mu}$, recoil and weak magnetism in
elastic nucleon scattering, elastic scattering on electrons
$\nu_{\mu}\re^\pm\to\re^\pm\nu_{\mu}$ and elastic scattering on
electron neutrinos and anti-neutrinos $\nu_{\mu}\nue\to \nue\nu_{\mu}$
and $\nu_{\mu}\bar\nue\to \bar\nue\nu_{\mu}$. The least important
processes are neutrino-neutrino scattering and $\re^+\re^-$
annihilation. The formation of the spectra and fluxes of $\nu_\mu$ is
dominated by the nucleonic processes, i.e.\ bremsstrahlung and elastic
scattering with recoil, but also $\nue\bar\nue$ annihilation and
$\nu_{\mu}\re^\pm$ scattering contribute significantly.  When all
processes are included, the spectral shape of the emitted neutrino
flux is always ``pinched,'' i.e.\ the width of the spectrum is smaller
than that of a thermal spectrum with the same average energy.  In all
of our cases we find that the average $\bar\nu_{\mu}$ energy exceeds
the average $\bar\nue$ energy by only a small amount, 10\% being a
typical number.  Weak magnetism effects cause the opacity of $\nu_\mu$
to differ slightly from that of $\bar\nu_\mu$, translating into
differences of the luminosities and average energies of a few percent.
Depending on the density, temperature, and composition profile, the
flavor-dependent luminosities $L_{\nue}$, $L_{\bar\nue}$, and
$L_{\nu_\mu}$ can mutually differ from each other by up to a factor of
two in either direction.
\end{abstract}



\section{\uppercase{Introduction}}

\label{sec:Introduction}

In numerical core-collapse supernova (SN) simulations, the transport
of $\mu$- and $\tau$-neutrinos has received scant attention because
their exact fluxes and spectra are probably not crucial for the
explosion mechanism.  However, the recent experimental evidence for
neutrino oscillations implies that the flavor-dependent fluxes and
spectra emitted by a SN will be partly swapped so that at any distance
from the source the actual fluxes and spectra can be very different
from those originally produced.  In principle, this effect can be
important for the SN shock revival (Fuller et~al.\ 1992) and r-process
nucleosynthesis (Qian et~al.\ 1993, Pastor \& Raffelt 2002), although
the experimentally favored small neutrino mass differences suggest
that this is not the case.  On the other hand, in view of the
large-mixing-angle solution of the solar neutrino problem flavor
oscillations are quite relevant for the interpretation of the SN~1987A
neutrino signal (Jegerlehner, Neubig, \& Raffelt 1996, Lunardini \&
Smirnov 2001a, Kachelriess et~al.\ 2002, Smirnov, Spergel, \& Bahcall
1994).  More importantly, the high-statistics neutrino signal from a
future galactic SN may allow one to differentiate between some of the
neutrino mixing scenarios which explain the presently available data
(Chiu \& Kuo 2000, Dighe \& Smirnov 2000, Dutta et~al.\ 2000, Fuller,
Haxton, \& McLaughlin 1999, Lunardini \& Smirnov 2001b, 2003, Minakata
\& Nunokawa 2001, Takahashi \& Sato 2002).  Even though the solution
of the solar neutrino problem has been established, the magnitude of
the small mixing angle $\Theta_{13}$ and the question if the neutrino
mass hierarchy is normal or inverted will remain open and can be
settled only by future precision measurements at dedicated
long-baseline oscillation experiments (Barger et~al.\ 2001, Cervera
et~al.\ 2000, Freund, Huber, \& Lindner 2001) and/or the observation
of a future galactic SN.

The usefulness of SN neutrinos for diagnosing flavor oscillations
depends on the flavor dependence of the fluxes and spectra at the
source.  Very crudely, a SN core is a black-body source of neutrinos
of all flavors which are emitted from the surface of the proto-neutron
star that was born after collapse.  It is the flavor-dependent details
of the neutrino transport in the neutron-star atmosphere which cause
the spectral and flux differences that can lead to interesting
oscillation effects.

The $\nue$ and $\bar\nue$ opacity is dominated by the charged-current
processes $\nue\rn\to\rp\re^-$ and $\bar\nue\rp\to\rn\re^+$, reactions
that allow for the exchange of energy and lepton number between the
medium and the neutrinos.  Therefore, it is straightforward to define
an energy-dependent neutrinosphere where this reaction freezes out for
neutrinos of a particular energy. This sphere yields a thermal
contribution to the neutrino flux at the considered energy.  The
atmosphere of the proto-neutron star is neutron rich, providing for a
larger $\nue$ opacity than for $\bar\nue$ so that for a given energy
the $\bar\nue$ flux originates at deeper and thus hotter layers than
the $\nue$ flux. In other words, a larger fraction of the $\bar\nue$
flux emerges with high energies.  This simple observation explains the
usual hierarchy $\langle \epsilon_{\bar\nue}\rangle>\langle
\epsilon_{\nue}\rangle$ of the mean energies.  The spectra are found
to be ``pinched'', meaning that the high-energy tail is suppressed
relative to that of a thermal spectrum with the same mean energy
(Janka \& Hillebrandt 1989a,b).  This numerical result can be
understood analytically by constructing the neutrino spectrum from the
fluxes emitted by the energy-dependent neutrinospheres which are at
different temperatures (Myra, Lattimer, \& Yahil 1988, Giovanoni,
Ellison, \& Bruenn 1989).

The formation of the $\nu_\mu$, $\bar\nu_\mu$, $\nu_\tau$, and
$\bar\nu_\tau$ spectra is far more complicated.  The opacity is
dominated by the neutral-current scattering on nucleons,
$\nu_\mu\rN\to\rN\nu_\mu$, a process that prevents neutrino free
streaming, but is unable to change the neutrino number and is usually
considered to be inefficient at exchanging energy.  (Here and in the
following $\nu_\mu$ stands for either $\nu_\mu$ or $\nu_\tau$.)
Neutrino pairs can be created by nucleon bremsstrahlung,
$\rN\rN\to\rN\rN \nu_\mu \bar\nu_\mu$, and pair annihilation,
$\re^-\re^+\to \nu_\mu\bar\nu_\mu$ or $\nue\bar\nue\to
\nu_\mu\bar\nu_\mu$, while $\nu_\mu\bar\nu_\mu$ pairs are absorbed by
the inverse reactions.  In addition, energy is exchanged by elastic
scattering on leptons, notably $\nu_\mu\re^-\to \re^-\nu_\mu$, by the
recoil in nucleon scattering, $\nu_\mu\rN\to\rN\nu_\mu$, and by
inelastic scattering on nucleons $\nu_\mu\rN\rN\to\rN\rN\nu_\mu$, a
channel that is the ``crossed process'' of bremsstrahlung.  For a
given neutrino energy these processes freeze out at different radii so
that one can define a ``number sphere'' for the pair processes, an
``energy sphere'' for the energy-exchange processes, and a ``transport
sphere'' for elastic nucleon scattering with $R_{\rm number}<R_{\rm
energy}<R_{\rm transport}$ (Suzuki 1990).  The region between the
number sphere and the transport sphere plays the role of a {\em
scattering atmosphere\/} because neutrinos can not be created or
destroyed. They propagate by diffusion and can still exchange energy
with the background medium.

Usually the $\nu_{\mu}$ transport sphere is deeper than the $\bar\nue$
sphere so that numerical simulations find
$\langle\epsilon_{\nu_\mu}\rangle>\langle\epsilon_{\bar\nue}\rangle
>\langle\epsilon_{\nue}\rangle$.  This hierarchy is the main
motivation for the proposed use of SN neutrinos as a diagnostic for
neutrino oscillations.  However, the quantitative statements found in
the literature range from $\langle\epsilon_{\nu_\mu}\rangle$ being
20\% to nearly a factor of 2 larger than
$\langle\epsilon_{\bar\nue}\rangle$; for a review see Janka (1993) and
Sec.~\ref{sec:PreviousLiterature}.  Of course, the mean energies and
their ratios change significantly between the SN bounce, accretion
phase, and the later neutron-star cooling phase.  Therefore, one must
distinguish carefully between instantaneous fluxes and spectra and the
time-integrated values.  While for the analysis of the sparse SN~1987A
data only time-integrated values make sense, a future galactic SN may
well produce enough events to study the instantaneous fluxes and
spectra (Barger, Marfatia, \& Wood 2001, Minakata et~al.\ 2001).

The overall energy emitted by a SN is often said to be equipartitioned
among all six neutrino degrees of freedom.  In some numerical
simulations the neutrino luminosities are indeed astonishingly equal
for all flavors (Totani et~al.\ 1998), while other simulations easily
find a factor of two difference between, say, the $\bar\nu_\mu$ and
$\bar\nue$ luminosities, at least during the accretion phase
(Mezzacappa et~al.\ 2001).  Therefore, it is by no means obvious how
precisely equipartition can be assumed for the purpose of diagnosing
neutrino oscillations.

Another important feature is the neutrino spectral shape, notably the
amount of pinching.  If one could assume with confidence that the
instantaneous spectra of all flavors are pinched at the source, and if
the measured SN neutrino spectra were instead found to be
anti-pinched, this effect would be a powerful diagnostic for the
partial spectral swapping caused by flavor oscillations (Dighe \&
Smirnov 2000).

Unfortunately, the existing literature does not allow one to develop a
clear view on these ``fine points'' of the neutrino fluxes and
spectra, largely because not enough attention has been paid to the
$\nu_\mu$ and $\nu_\tau$ emission from a SN core.  The published full
numerical SN collapse simulations have not yet included the
bremsstrahlung process or nucleon recoils (but see first results of
state-of-the-art models in Rampp et al.\ 2002), even though it is no
longer controversial that these effects are important (Janka et~al.\
1996, Burrows et~al.\ 2000, Hannestad \& Raffelt 1998, Raffelt 2001,
Suzuki 1991, 1993, Thompson, Burrows, \& Horvath 2000).  Moreover,
some of the interesting information such as the spectral pinching was
usually not documented.

Another problem with self-consistent hydrodynamic simulations is
that the models with the most elaborate neutrino transport usually do
not explode so that even the most recent state-of-the-art simulations
do not reach beyond the accretion phase at a few hundred milliseconds
after bounce (Rampp \& Janka 2000, Mezzacappa et~al.\ 2001, 
Liebend\"orfer et~al.\ 2001), thus not
providing any information on the neutron-star cooling phase.
Successful multi-dimensional models of the explosion (e.g., Fryer \&
Warren 2002, Fryer 1999 and references therein) were also not
continued to the neutron-star cooling phase. These simulations,
moreover, treat the neutrino transport only in a very approximate way
and do not provide spectral information. The calculations performed by
the Livermore group also yield robust explosions (Totani et~al.\
1998).  They include a mixing-length treatment of the phenomenon of
neutron-finger convection in the neutron star, that increases the
early neutrino luminosities and thus enhances the energy transfer by
neutrinos to the postshock medium (Wilson \& Mayle 1993). Whether
neutron-finger convection actually occurs inside the neutrinosphere
and has effects on a macroscopic scale, however, is an unsettled
issue.

We will follow here an alternative approach to full hydrodynamic
simulations, i.e.\ we will study neutrino transport on the background
of an assumed neutron-star atmosphere.  While this approach lacks
hydrodynamic self-consistency, it has the great advantage of
allowing one to study systematically the influence of various pieces
of microscopic input physics and of the medium profile.  The goal is
to develop a clearer picture of the generic properties of the SN
neutrino spectra and fluxes and what they depend upon.

To this end we have adapted the Monte Carlo code of Janka (1987, 1991)
and added new microphysics to it. We go beyond the work of Janka \&
Hillebrandt (1989a,b) in that we include the bremsstrahlung process,
nucleon recoils and weak magnetism, $\nue \bar\nue$ pair annihilation into
$\nu_\mu\bar\nu_\mu$, and scattering of $\nu_{\mu}$ on $\nue$ and
$\bar\nue$. With these extensions we investigate the neutrino
transport systematically for a variety of medium profiles that are
representative for different SN phases.  One of us (Raffelt 2001) has
recently studied the $\nu_\mu$ spectra-formation problem with the
limitation to nucleonic processes (elastic and inelastic scattering,
recoils, bremsstrahlung), to Maxwell-Boltzmann statistics for the
neutrinos, and plane-parallel geometry.  Our present study complements
this more schematic work by including the leptonic processes,
Fermi-Dirac statistics, and spherical geometry.  In addition we apply
our Monte Carlo code to the transport of $\nue$ and $\bar\nue$ and
thus are able to compare the flavor-dependent fluxes and spectra.

In Sec.~2 we first assess the relative importance of different
processes in terms of their energy-dependent ``thermalization depth''.
In this context we introduce a number of stellar background models.
In Sec.~3 we perform a Monte Carlo study of $\nu_\mu$ transport on the
previously introduced background models in order to assess the
importance of different pieces of input physics.  In Sec.~4 we compare
the $\nu_\mu$ fluxes and spectra with those of $\nue$ and
$\bar\nue$. We conclude in Sec.~5 with a discussion and summary of our
findings.


\section{\uppercase{Thermalization Depth of\\
        Energy-Exchange Processes}}


\subsection{Simple Picture of Spectra Formation}

\label{sec:SimplePicture}

One of our goals is to assess the relative importance of different
neutrino interaction channels with the background medium of the SN
core.  As a first step it is instructive to study the thermalization
depth of various energy-exchange processes.  Within the transport
sphere, the neutrinos are trapped by elastic scatterings on nucleons,
$\nu_\mu\rN\to\rN\nu_\mu$, which are by far the most frequent
reactions between neutrinos and particles of the stellar medium.
(Unless otherwise noted ``neutrinos'' always refers to any of
$\nu_\mu$, $\bar\nu_\mu$, $\nu_\tau$ or $\bar\nu_\tau$.) Assuming for
the moment that these collisions are iso-energetic (no nucleon
recoils), it is straightforward to define for a neutrino of given
energy~$\epsilon$ the location (``thermalization depth'') where it
last exchanged energy with the medium by a reaction such as
$\nu_\mu\re^-\to\re^-\nu_\mu$.  Following Shapiro \& Teukolsky (1983)
we define the optical depth for energy exchange or thermalization by
\begin{equation}\label{eq:definetautherm}
\tau_{\rm therm}(r)
=\int_r^\infty \!\! dr'\sqrt{\frac{1}{\lambda_E(r')}
\left[\frac{1}{\lambda_T(r')}+\frac{1}{\lambda_E(r')}\right]}\,.
\end{equation} 
Here, $\lambda_E$ is the mean free path (mfp) for the relevant
energy-exchange process and $\lambda_T$ the transport mfp, i.e.\ the
mfp corresponding to the cross section for momentum exchange in the
$\nu\rN\to\rN\nu$ reaction.  The quantities $\tau_{\rm therm}$,
$\lambda_E$ and $\lambda_T$ are all understood to depend on the
neutrino energy~$\epsilon$.  The main philosophy of
Eq.~(\ref{eq:definetautherm}) is that a neutrino trapped by elastic
scattering has a chance to exchange energy corresponding to its actual
diffusive path through the scattering atmosphere; for a discussion see
Suzuki (1990). The thermalization depth $R_{\rm therm}$ is given by
\begin{equation}\label{eq:defineRtherm}
\tau_{\rm therm}(R_{\rm therm})=\frac{2}{3}\,,
\end{equation}
where $R_{\rm therm}$ depends on the neutrino energy $\epsilon$.

When this energy dependence is not too steep it makes sense to define
an average thermalization depth, i.e.\ an ``energy sphere'' that for
pair creating processes is equal to the ``number sphere.''  For
nucleon bremsstrahlung this requirement is well fulfilled (Raffelt
2001) so that one may picture the energy sphere as a blackbody surface
that injects neutrinos into the scattering atmosphere and absorbs
those scattered back.  The neutrino flux and spectrum emerging from
the transport sphere is then easily understood in terms of the
energy-dependent transmission probability of the blackbody spectrum
launched at the energy sphere.  The transport cross section scales as
$\epsilon^2$, implying that the transmitted flux spectrum is shifted
to lower energies relative to the temperature at the energy sphere.
This simple ``filter effect'' accounts surprisingly well for the
emerging flux spectrum (Raffelt 2001).  For typical conditions the
mean flux energies are 50--60\% of those corresponding to the
blackbody conditions at the energy sphere.

Moreover, it is straightforward to understand that the effective
temperature of the emerging flux spectrum is not overly sensitive to
the exact location of the energy sphere.  If the energy-exchange
reaction is somewhat more effective, the energy sphere is at a larger
radius with a lower medium temperature.  However, the scattering
atmosphere has a smaller optical depth so that the higher-energy
neutrinos are less suppressed by the filter effect, partly
compensating the smaller energy-sphere temperature.  For typical
situations Raffelt (2001) found that changing the bremsstrahlung rate
by a factor of 3 would change the emerging neutrino energies only by
some 10\%.  This finding suggests that the emitted average neutrino
energy is not overly sensitive to the details of the energy-exchange
processes.


\subsection{Neutron-Star Atmospheres}
\label{sec:NeutronStarAtmospheres}

\begin{deluxetable}{lll}
\tablecaption{\label{tab:powerlawmodels} 
Characteristics of power-law models.}
\tablewidth{0pt}
\tablehead{&Steep&Shallow}
\startdata
$p$                              & 10    & 5   \\
$q$                              & 2.5   & 1   \\
$q/p$                            & 0.25  & 0.2 \\
$\rho_0~\rm [10^{14}~g\,cm^{-3}]$ & 2.0   & 0.2 \\
$T_0~\rm [MeV]$                  & 31.66 & 20.0\\
$r_0~\rm [km]$                   & 10    & 10  \\
\enddata
\end{deluxetable}

\begin{figure}[b]
\columnwidth=6.5cm
\plotone{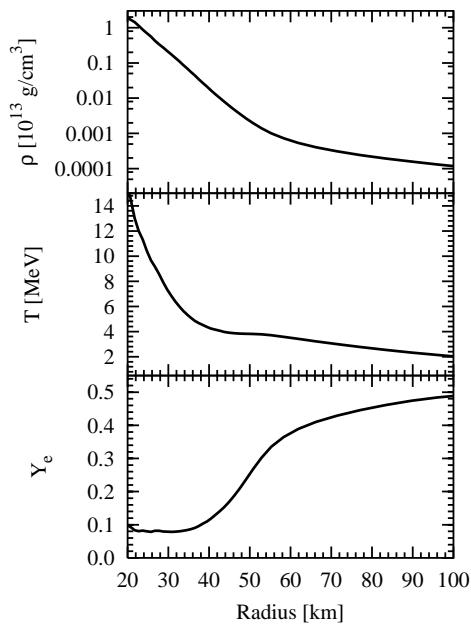}
\caption{\label{fig:Messermodel} Accretion-Phase Model~I, a SN model
324~ms after bounce from a Newtonian calculation (O.E.B.~Messer,
personal communication).}
\end{figure}

\begin{figure}[b]
\columnwidth=6.5cm
\plotone{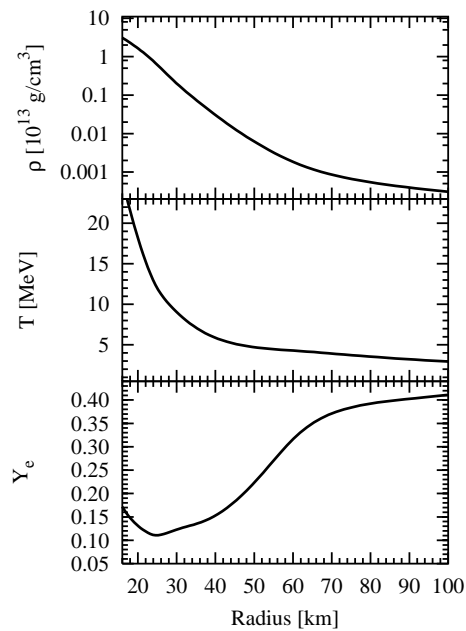}
\caption{\label{fig:Ramppmodel} Accretion-Phase Model~II, a SN core at
150~ms postbounce from a general-relativistic simulation.  (M.~Rampp,
personal communication).}
\end{figure}

In order to determine the location of the thermalization depth for
different processes we need to define our assumed neutron-star
atmospheres.  As a first example we use a model taken from a full
hydrodynamic simulation. This model is representative for the
accretion phase; henceforth we will refer to it as the
``Accretion-Phase Model~I'' (Fig.~\ref{fig:Messermodel}).  It was
provided to us by O.~E.~B.~Messer and was already used in Raffelt
(2001) for a more schematic study.  Based on the Woosley \& Weaver
$15\,M_\odot$ progenitor model labeled s15s7b, the Newtonian collapse
simulation was performed with the SN code developed by Mezzacappa
et~al.\ (2001).  The snapshot is taken at 324~ms after bounce when the
shock is at about 120~km, i.e.\ the star still accretes matter.  In
this simulation the traditional microphysics for $\nu_\mu$ transport
was included, i.e.\ iso-energetic scattering on nucleons, $\re^+\re^-$
annihilation and $\nu_\mu\re^-$ scattering.

As another self-consistent example (Accretion-Phase Model~II) we
obtained a 150~ms postbounce model from M.~Rampp (personal
communication) that uses a very similar progenitor (s15s7b2).  The
simulation includes an approximate general relativistic treatment in
spherical symmetry as described by Rampp \& Janka (2002).  The three
neutrino flavors are transported with all relevant interactions except
$\nue\bar\nue$ pair annihilation to $\nu_{\mu}\bar\nu_{\mu}$ (see also
Sec.~\ref{sec:ourflavorcomp} and Rampp et al.\ 2002).

As another set of examples we use two power-law profiles of the form
\begin{equation}\label{eq:powerlaws}
\rho=\rho_0\left(\frac{r_0}{r}\right)^p,\quad
T=T_0\left(\frac{r_0}{r}\right)^q,
\end{equation}
with a constant electron fraction per baryon $\Ye$.  We adjust
parameters such that $\langle\epsilon\rangle\approx 20$--25 MeV for
the emerging neutrinos to obtain model atmospheres in the ballpark of
results from proto-neutron star evolution calculations.  We define a
``steep'' power-law model, corresponding to the one used by Raffelt
(2001), and a ``shallow'' one; the characteristics are given in
Table~\ref{tab:powerlawmodels}.  The shallow model could be
characteristic of a SN core during the accretion phase while the steep
model is more characteristic for the neutron-star cooling phase. The
constant electron fraction $\Ye$ is another parameter that allows us
to investigate the relative importance of the leptonic processes as a
function of the assumed $\Ye$.


\subsection{Thermalization Depth}

\begin{figure}[t]
\columnwidth=6.5cm
\plotone{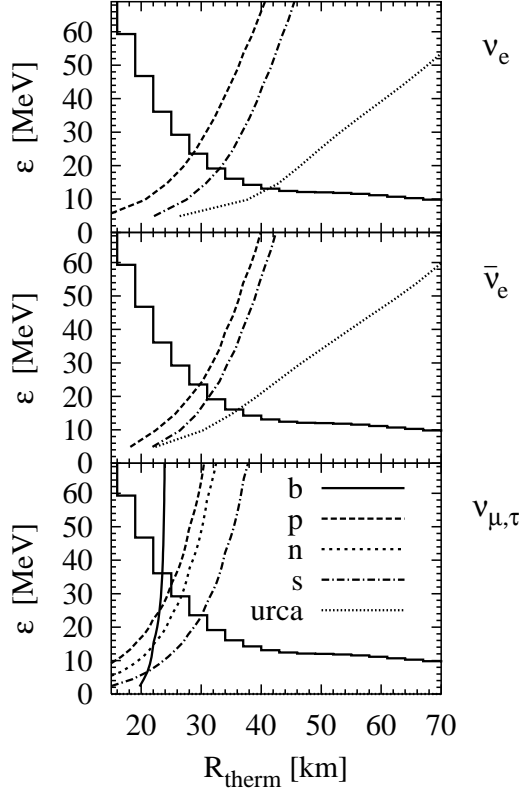}
\caption{\label{fig:rthermrealistic} $R_{\rm therm}$ as a function of
neutrino energy $\epsilon$ for our Accretion-Phase Model~I.  From top
to bottom the panels show the results for $\nue$, $\bar\nue$, and
$\nu_\mu$.  Energy exchanging processes: bremsstrahlung (solid line),
$\re^+\re^-$ annihilation (dashed), $\nue \bar\nue$ annihilation
(dotted), and scattering on $\re^\pm$ (dash-dotted). ``Urca'' denotes
the charged-current reaction of $\nue$ and $\bar\nue$ on nucleons.
The steps represent $\langle\epsilon\rangle = 3.15\,T$.}
\end{figure}

\begin{figure}[t]
\columnwidth=6.5cm
\plotone{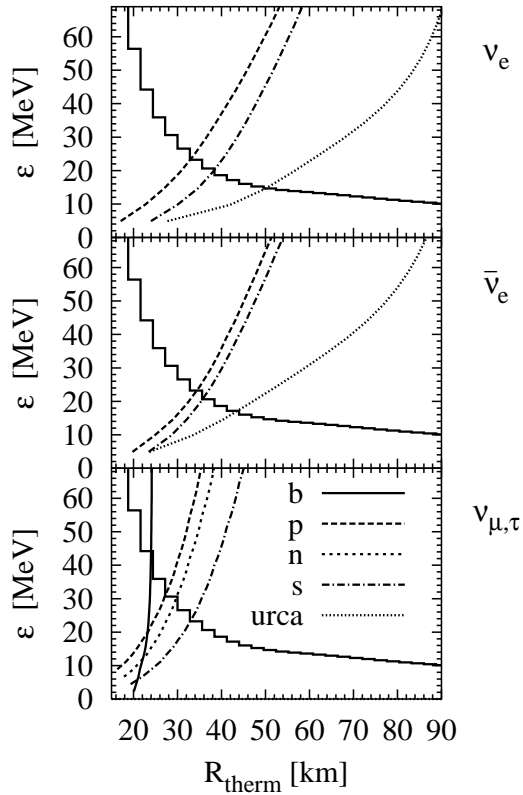}
\caption{\label{fig:rthermrealistic1} Same as
Fig.~\ref{fig:rthermrealistic} for the Accretion-Phase Model~II.}
\end{figure}

\begin{figure}[t]
\columnwidth=6.5cm
\plotone{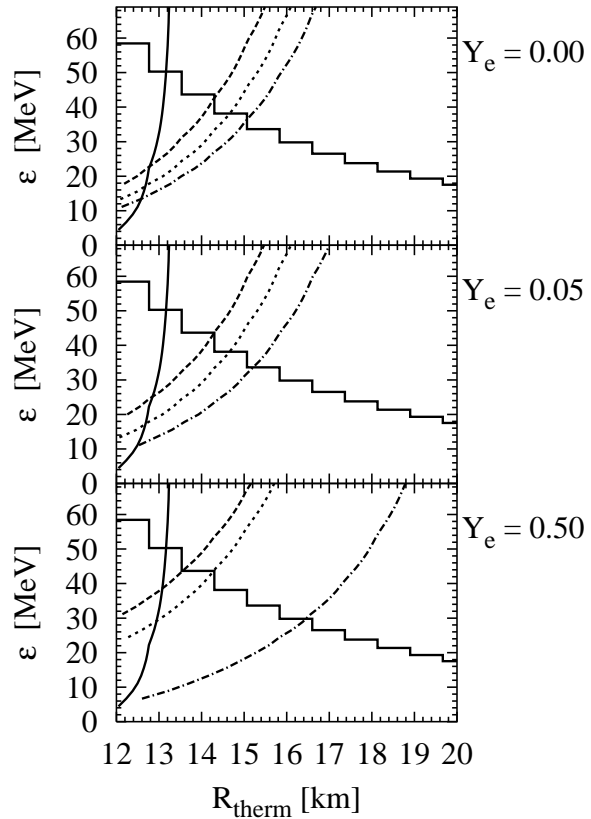}
\caption{\label{fig:rthermsteep} $R_{\rm therm}$ for $\nu_\mu$ in the
steep power-law model with the indicated values of $\Ye$. This figure
corresponds to the bottom panel of Fig.~\ref{fig:rthermrealistic}.}
\end{figure}

\begin{figure}[t]
\columnwidth=6.5cm
\plotone{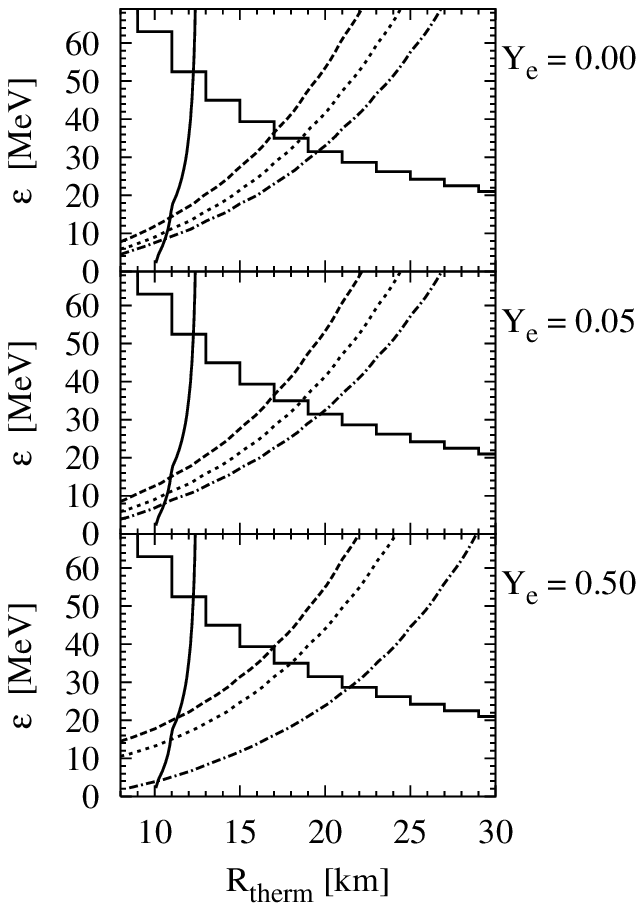}
\caption{\label{fig:rthermshallow}
 Same as Fig.~\ref{fig:rthermsteep} for the shallow 
power-law model.}
\end{figure}

We now calculate the thermalization depth as a function of the
neutrino energy $\epsilon$ for several energy-exchanging processes and
the neutron-star atmospheres described above.  We consider the
neutrino mfp for nucleon bremsstrahlung
$\rN\rN\to\rN\rN\nu_\mu\bar\nu_\mu$, pair annihilation $\re^+\re^-\to
\nu_\mu\bar\nu_\mu$ and $\nue\bar\nue \to \nu_\mu\bar\nu_\mu$, and
scattering on charged leptons $\nu_\mu \re^\pm\to\re^\pm\nu_\mu$.  The
numerical implementation of the reaction rates is described in
Appendix~\ref{appendix:rates}.

In Figs.~\ref{fig:rthermrealistic} and \ref{fig:rthermrealistic1} we
give the thermalization depth $R_{\rm therm}$ as a function of
neutrino energy $\epsilon$ for the two hydrodynamically
self-consistent accretion-phase models.  From top to bottom the panels
show the results for $\nue$, $\bar\nue$, and $\nu_\mu$, respectively.
The step-like curves represent the temperature profiles in terms of
the mean neutrino energy, $\langle\epsilon_{\nu}\rangle=3.15\,T$ for
non-degenerate neutrinos at the local medium temperature; the steps
correspond to the radial zones of our Monte Carlo simulation.  The
other curves represent $R_{\rm therm}$ for bremsstrahlung (b), $\re^+
\re^-$ annihilation (p), $\nue\bar\nue$ annihilation (n), and
scattering on $\re^\pm$ (s).  In the case of $\nue$ and $\bar\nue$ we
do not include bremsstrahlung and $\nue\bar\nue$ annihilation.
Particle creation is dominated by the charged current reactions on
nucleons (urca).

For the power-law models we show $R_{\rm therm}$ for $\nu_\mu$ in
Figs.~\ref{fig:rthermsteep} and~\ref{fig:rthermshallow}.  The
different panels correspond to the indicated values of the electron
fraction $\Ye$.  Note that $\Ye$ represents the net electron density
per baryon, i.e.\ the $\re^-$ density minus that of $\re^+$ so that
$\Ye=0$ implies that there is an equal thermal population of $\re^-$
and $\re^+$.

The $\nu_\mu$ absorption rate for the bremsstrahlung process varies
approximately as $\epsilon^{-1}$, the $\nu_\mu\rN$ transport cross
section as $\epsilon^2$ so that the inverse mfp for thermalization
varies only as $\epsilon^{1/2}$.  This explains why $R_{\rm therm}$
for bremsstrahlung is indeed quite independent of $\epsilon$.
Therefore, bremsstrahlung alone allows one to specify a rather
well-defined energy sphere.  The other processes depend much more
sensitively on $\epsilon$ so that a mean energy sphere is much
less well defined.

Both electron scattering and the leptonic pair processes are so
ineffective at low energies that true local thermodynamic equilibrium
(LTE) can not be established even for astonishingly deep locations.
Bremsstrahlung easily ``plugs'' this low-energy hole so that one can
indeed expect LTE for all relevant neutrino energies below a certain
radius.  For higher energies, the leptonic processes dominate and
shift the energy sphere to larger radii than bremsstrahlung alone.
The relative importance of the various processes depends on the
density and temperature profiles as well as $\Ye$.

To assess the role of the various processes for the overall spectra
formation one needs to specify some typical neutrino energy.  One
possibility would be $\langle\epsilon\rangle$ for neutrinos in LTE.
Another possibility is the mean energy of the neutrino flux, in
particular the mean energy of those neutrinos which actually leave the
star.  For our power-law atmospheres this is always around 20--25~MeV.
Therefore, the process with the largest $R_{\rm therm}$ in this energy
band is the one most relevant for determining the emerging neutrino
spectrum.  It appears that at least for steep profiles pair
annihilation is never crucial once bremsstrahlung is included, i.e.\
we would guess that including pair annihilation will not affect the
emerging neutrino spectra.  The relevance of electron scattering is
far more difficult to guess. On the one hand it surely is more
important than recoil in nucleon scatterings for some of the relevant
energies, on the other hand we are not able to define an energy sphere
for nucleon recoils because this process is different from the others
in that neutrinos transfer only a small fraction of their energy per
scattering.  Therefore, it is not straightforward to assess the
relevance of electron scattering compared with nucleon recoils on the
basis of the various thermalization spheres alone.


\section{\uppercase{Monte Carlo Study of Muon Neutrino Transport}}


\subsection{Spectral Characteristics}
\label{sec:specchar}

In order to characterize the neutrino spectra and fluxes emerging from
a neutron star we need to introduce some simple and intuitive
parameters.  One is the mean energy
\begin{equation}\label{eq:meanenergy}
\langle\epsilon\rangle=
\frac{\int_0^\infty d\epsilon\,\epsilon\,\int_{-1}^{+1}d\mu\,
f(\epsilon,\mu)}
{\int_0^\infty d\epsilon\,\int_{-1}^{+1}d\mu\,f(\epsilon,\mu)}\,,
\end{equation} 
where $f(\epsilon,\mu)$ is the neutrino distribution function with
$\epsilon$ the energy and $\mu$ the cosine of the angle between the
neutrino momentum and the radial direction.  If the neutrinos
are in LTE without a chemical potential one has
\begin{equation}\label{eq:eqdist}
f(\epsilon,\mu) = \frac{\epsilon^2}{1+\exp(\epsilon/T)}
\end{equation}
and therefore
\begin{equation}\label{eq:avenergytot}
\langle\epsilon\rangle=\frac{7\pi^4}{180\,\zeta_3}\,T
\approx3.1514\,T\,.
\end{equation}
One can define an effective neutrino temperature for non-equilibrium
distributions by inverting this relationship.

It is often useful to extract spectral characteristics for those
neutrinos which are actually flowing by removing the isotropic part of
the distribution.  Specifically, we define the average flux energy by
\begin{equation}
\label{eq:fluxenergy}
\langle\epsilon\rangle_{\rm flux}=
\frac{\int_0^\infty d\epsilon\,\epsilon\,\int_{-1}^{+1}d\mu\,\mu\,
f(\epsilon,\mu)}
{\int_0^\infty d\epsilon\,\int_{-1}^{+1}d\mu\,\mu\,f(\epsilon,\mu)}\,.
\end{equation} 
Far away from the star all neutrinos will flow essentially in the
radial direction, implying that the angular distribution becomes a
delta-function in the forward direction so that
$\langle\epsilon\rangle_{\rm flux}=\langle\epsilon\rangle$.  However,
in the trapping regions the two averages are very different because
the distribution function is dominated by its isotropic term.

To characterize the spectrum beyond the mean energy one can
consider a series of moments $\langle\epsilon^n\rangle$ (Janka \&
Hillebrandt 1989a); we usually limit ourselves to $n=1$ and~2.  Note
that a Fermi-Dirac distribution at zero chemical potential yields
\begin{equation}
a\equiv\frac{\langle\epsilon^2\rangle}{\langle\epsilon\rangle^2}
=\frac{486000\,\zeta_3\zeta_5}{49\,\pi^8}\approx1.3029\,.
\end{equation}
For a Maxwell-Boltzmann distribution this quantity would be 4/3. 
Following Raffelt (2001) we further define the ``pinching parameter''
\begin{equation}
p\equiv \frac{1}{a}\,
\frac{\langle\epsilon^2\rangle}{\langle\epsilon\rangle^2}\,,
\end{equation}
where $p=1$ signifies that the spectrum is thermal up to its second
moment, while $p<1$ signifies a pinched spectrum (high-energy tail
suppressed), $p>1$ an anti-pinched spectrum (high-energy tail
enhanced).  An analogous definition applies to the pinching parameter
$p_{\rm flux}$ of the flux spectrum by replacing $\langle\cdot\rangle$
with $\langle\cdot\rangle_{\rm flux}$.

In some publications the root-mean-square energy
$\langle\epsilon\rangle_{\rm rms}$ is given instead of the average
energy.  The definition corresponding to Eq.~(\ref{eq:meanenergy}) is
\begin{equation}
\label{eq:rmsdef}
\langle\epsilon\rangle_{\rm rms}=\sqrt{
{\int_0^\infty\!d\epsilon\int_{-1}^{+1}\!d\mu\,
\epsilon^3f(\epsilon,\mu) \over
\int_0^\infty\!d\epsilon\int_{-1}^{+1}\!d\mu\,
\epsilon\,f(\epsilon,\mu)} } =
\sqrt{\frac{\langle\epsilon^3\rangle}{\langle\epsilon\rangle}}\,.
\end{equation}
This characteristic spectral energy is useful for estimating 
the energy transfer from neutrinos to the stellar medium in
reactions with cross sections proportional to $\epsilon^2$.
For thermal neutrinos with vanishing chemical potential we find
\begin{equation}
\langle\epsilon\rangle_{\rm rms}=\sqrt{\frac{930}{441}}\,\pi\,T
\approx 4.5622\,T\,.
\end{equation}
With Eq.~(\ref{eq:avenergytot}) this corresponds to
$\langle\epsilon\rangle \approx 0.691\langle\epsilon\rangle_{\rm
rms}$.

\begin{figure}[ht]
\columnwidth=6.5cm
\plotone{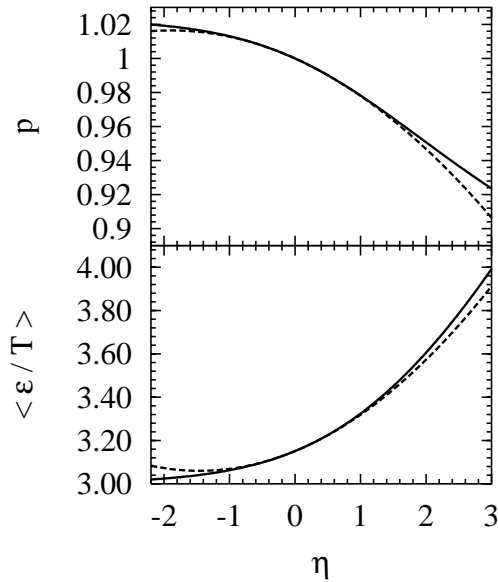}
\caption{\label{fig:fermidirac} Mean energy and pinching parameter
as a function of the degeneracy parameter for a Fermi-Dirac
distribution.  As dashed lines we show the expansions given in
Eq.~(\ref{eq:fdexpansion}).}
\end{figure}

Beyond the energy moments $\langle\epsilon^n\rangle$ and related
parameters it is often useful to approximate the neutrino spectrum by
a simple analytic fit. If one uses two parameters beyond the overall
normalization one can adjust the fit to reproduce two moments, for
example $\langle\epsilon\rangle$ and $\langle\epsilon^2\rangle$.  In
the literature one frequently encounters an approximation in terms of
a nominal Fermi-Dirac distribution characterized by a temperature $T$
and a degeneracy parameter $\eta$ according to
\begin{equation}\label{eq:FermiDiracfit}
f_\eta(\epsilon)=
\frac{\epsilon^2}{1+\exp\left(\frac{\epsilon}{T}-\eta\right)}
\end{equation}
(Janka \& Hillebrandt 1989a). In Fig.~\ref{fig:fermidirac} we show
$\langle\epsilon/T\rangle$ and $p$ as a function of $\eta$.
Up to second order, expansions are
\begin{eqnarray}\label{eq:fdexpansion}
\langle\epsilon/T\rangle&\approx& 3.1514+0.1250\,\eta+0.0429\,\eta^2
\nonumber\\
p&\approx& 1 - 0.0174\,\eta -0.0046\,\eta^2\,.
\end{eqnarray}
These expansions are shown in Fig.~\ref{fig:fermidirac} as dashed
lines.

Using a nominal Fermi-Dirac fit to approximate the spectrum is
physically motivated because a truly thermal neutrino flux would
follow this behavior. On the other hand, the neutrino flux emitted
from a SN core is not very close to being thermal so that the limiting
behavior of the fit function is not a strong argument. Therefore, we
consider an alternative fit function for which analytic simplicity is
the main motivation,
\begin{equation}\label{eq:powerlawfit}
f_\alpha(\epsilon)=\left(\frac{\epsilon}{\bar\epsilon}\right)^\alpha 
{\rm e}^{-(\alpha+1) \epsilon/\bar\epsilon}\,.
\end{equation} 
For any value of $\alpha$ we have
$\langle\epsilon\rangle=\bar\epsilon$ while
\begin{equation}
\label{eq:pinchingalpharelation}
\frac{\langle\epsilon^2\rangle}{\langle\epsilon\rangle^2}
=\frac{2+\alpha}{1+\alpha}\,.
\end{equation}
Put another way, $\bar\epsilon$ is the average energy while $\alpha$
represents the amount of spectral pinching.  For general moments the
analogous relation is \be
\label{eq:momentsalpharelation}
\frac{\langle\epsilon^k\rangle}{\langle\epsilon^{k-1}\rangle}
= \frac{k+\alpha}{1+\alpha} \;\bar\epsilon \,.
\ee

\begin{deluxetable}{lll}
\tablecaption{\label{tab:fitfunctions} 
Parameters for fit-functions of Fig.~\ref{fig:fitfunctions}.}
\tablewidth{0pt}
\tablehead{Width&$\alpha$&$\eta$}
\startdata
$w_0=\langle\epsilon\rangle/\sqrt{3}$&2.    &$-\infty$\\
$0.9\,w_0$                           &2.7037&1.1340\\
$0.8\,w_0$                           &3.6875&2.7054\\
$0.7\,w_0$                           &5.1225&4.4014\\
$0.6\,w_0$                           &7.3333&6.9691\\
$0.5\,w_0$                           &11.   &13.892\\
\enddata
\end{deluxetable}

\begin{figure}[ht]
\columnwidth=6.5cm
\plotone{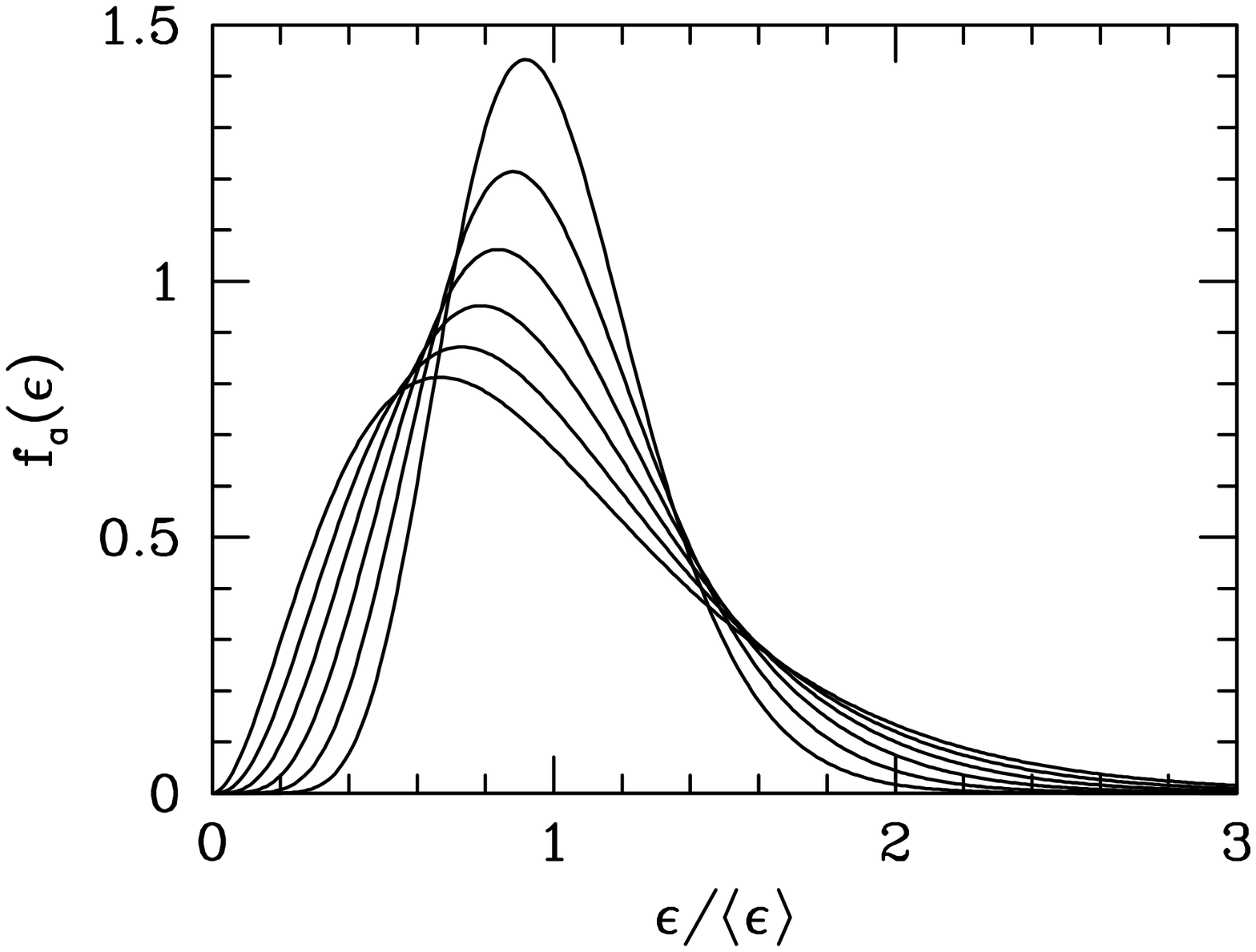}
\plotone{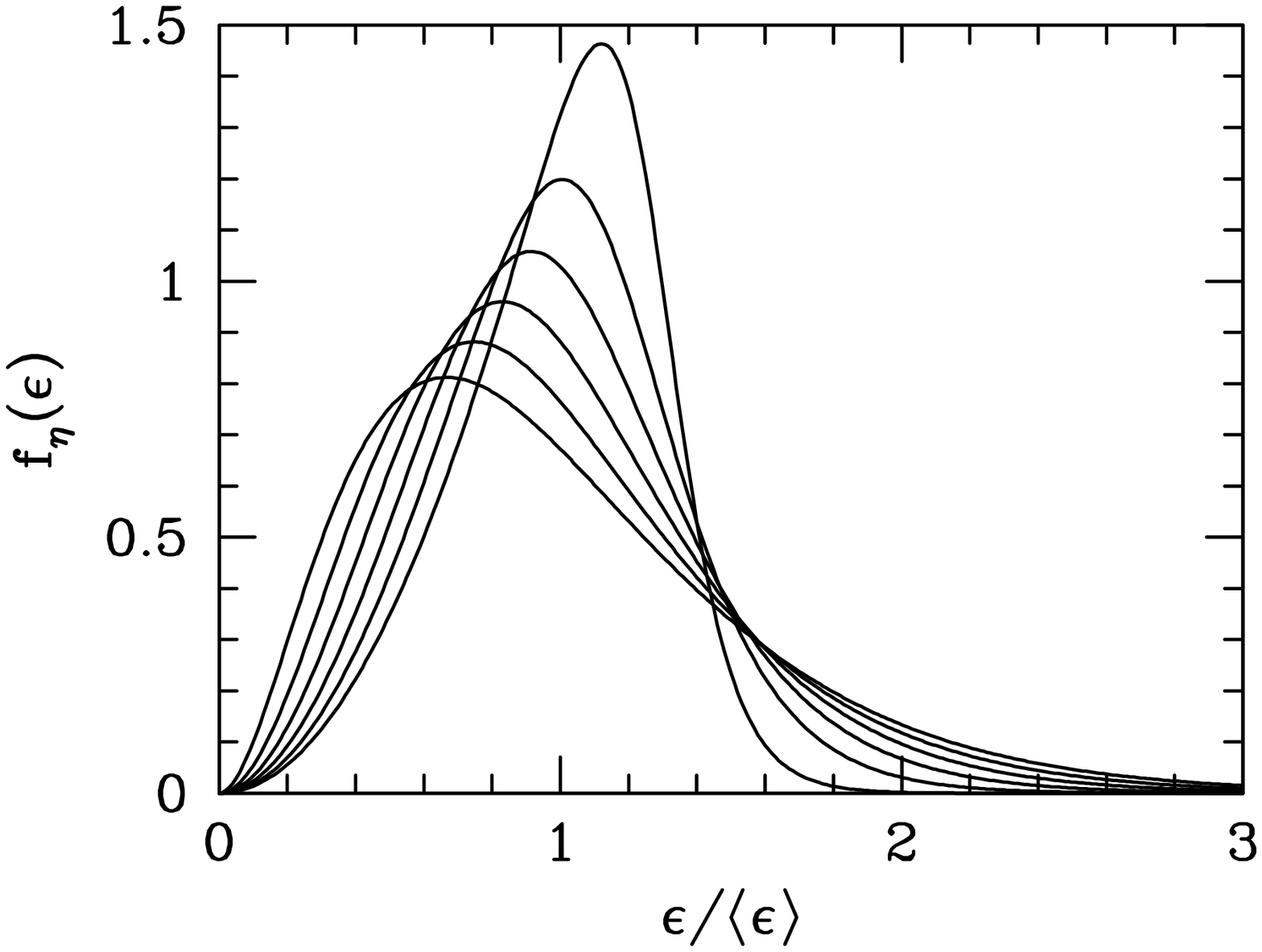}
\plotone{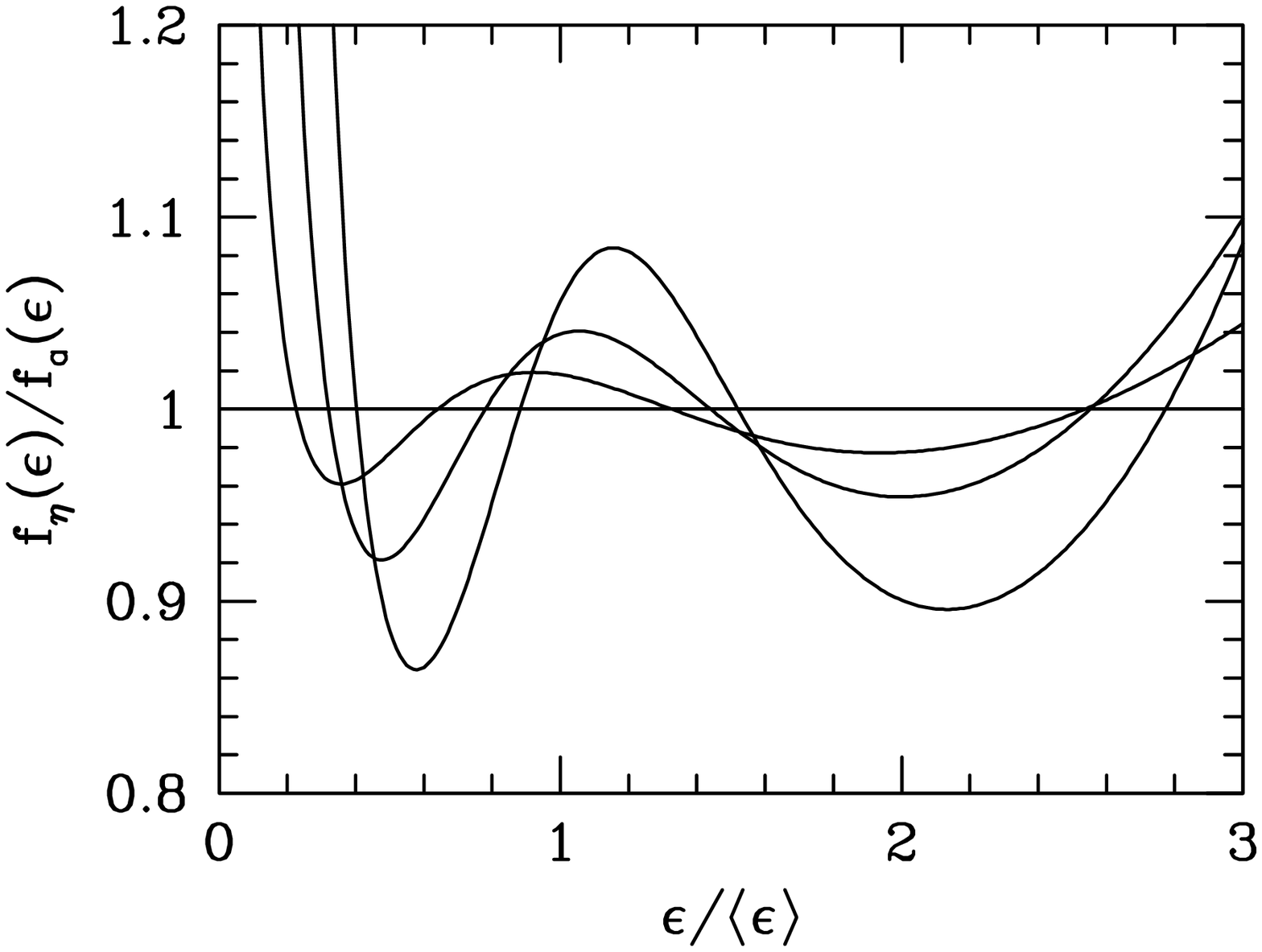}
\caption{\label{fig:fitfunctions} Normalized fit functions.  {\it
Upper panel:} ``Power law'' according to Eq.~(\ref{eq:powerlawfit}).
{\it Middle panel:} Fermi-Dirac fit according to
Eq.~(\ref{eq:FermiDiracfit}).  In both panels the broadest curve
corresponds to $f(\epsilon)=\epsilon^2 \exp(-3\epsilon/\bar\epsilon)$,
i.e.\ to $\alpha=2$ and $\eta=-\infty$, respectively.  For the other
curves the width was decreased in decrements of 10\%, see
Table~\ref{tab:fitfunctions}.  {\it Bottom panel:} Ratio of the fits
$f_\alpha/f_\eta$ for the first four cases.}
\end{figure}

In the upper panel of Fig.~\ref{fig:fitfunctions} we show
$f_\alpha(\epsilon)$, the integral normalized to unity, for several
values of $\alpha$. The broadest curve is for $\alpha=2$ while for the
narrower ones the width
\begin{equation}
w=\sqrt{\langle\epsilon^2\rangle-\langle\epsilon\rangle^2}
\end{equation}
was decreased in 10\% decrements as shown in
Table~\ref{tab:fitfunctions}. The middle panel of 
Fig.~\ref{fig:fitfunctions} shows the corresponding
curves $f_\eta(\epsilon)$ with the $\eta$-values given
in Table~\ref{tab:fitfunctions}. The broadest curves in each
panel are identical and correspond to
$\epsilon^2\exp(-3\epsilon/\bar\epsilon)$
with a width $w_0=\langle\epsilon\rangle/\sqrt{3}$.

The limiting behavior of $f_\alpha(\epsilon)$ for large $\alpha$ is
$\delta(\epsilon-\bar\epsilon)$ while for $f_\eta(\epsilon)$ the
limiting width is $w_0/\sqrt{5}\approx0.44721\,w_0$. Evidently the
curves $f_\alpha(\epsilon)$ can accommodate a much broader range of
widths than the curves $f_\eta(\epsilon)$.

We will find that the neutrino spectra are always fit with parameters
in the range $2\lsim\alpha\lsim5$ or $0\lsim\eta\lsim4$, i.e.\ with a
width above about $0.75\,w_0$. In the bottom panel of
Fig.~\ref{fig:fitfunctions} we show the ratios of the fit functions
for the widths down to $0.7\,w_0$. Except for the lowest energies and
very high energies the two fit functions are equivalent to better than
10\%. Therefore, the two types of fits are largely equivalent for most
practical purposes.

On the basis of a few high-statistics Monte-Carlo runs we will show in
Sec.~\ref{sec:High-Statistics} that the numerical spectra are actually
better approximated over a broader range of energies by the
``power-law'' fit functions $f_\alpha(\epsilon)$.  In addition, these
functions are more flexible at representing the high-energy tail of
the spectrum that is most relevant for studying the Earth effect in
neutrino oscillations.


\subsection{Monte Carlo Set Up}
\label{sec:MCsetup}

We have run our Monte Carlo code, that is described in
Appendix~\ref{app:MonteCarlo}, for the stellar background models
introduced in Sec.~\ref{sec:NeutronStarAtmospheres} and for different
combinations of energy-changing neutrino reactions.  Our main interest
is to assess the impact of the scattering atmosphere on the flux and
spectrum formation. Therefore, it is sufficient to simulate the
neutrino transport above some radius where we have to specify a
boundary condition.

We always use a blackbody boundary condition at the bottom of the
atmosphere, i.e.\ we assume neutrinos to be in LTE at the local
temperature and the appropriate chemical potential; for $\nu_\mu$ and
$\nu_\tau$ the latter is taken to vanish.  As a consequence of this
boundary condition, the luminosity emerging at the surface is
generated within the computational domain and calculated by our Monte
Carlo transport.  A small flux across the inner boundary develops
because of the negative gradients of temperature and density in the
atmosphere, but its magnitude depends on the radial resolution of the
neutron-star atmosphere and will not in general correspond to the
physical diffusive flux.  But as long as the flux is small compared to
the luminosity at the surface, the emerging neutrino spectra will not
depend on the lower boundary condition. Usually it is sufficient to
place the inner grid radius deeper in the star than the thermalization
depth of the dominant pair process.

The shallow energy dependence of the thermalization depth of the
nucleon bremsstrahlung implies that whenever we include this process
it is not difficult to choose a reasonable location for the lower
boundary.  Taking the latter too deep in the star is very
CPU-expensive as one spends most of the simulation for calculating
frequent scatterings of neutrinos that are essentially in LTE.

We always include $\nu_\mu\rN$ scattering as the main opacity source.
For energy exchange, we switch on or off bremsstrahlung~(b), nucleon
recoil~(r), scattering on electrons~(s), $\re^+ \re^-$ pair
annihilation (p), and $\nue \bar\nue$ annihilation~(n).  We never
include inelastic nucleon scattering $\nu_\mu\rN\rN\to\rN\rN\nu_\mu$
as this process is never important relative to recoil (Raffelt 2001).
Likewise, we ignore scattering on $\nue$ and $\bar\nue$ which is
always unimportant if $\nu_\mu\re^\pm$ is included (Buras et~al.\
2002).  We also neglect $\nu_\mu\nu_\mu$ or $\nu_\mu\bar\nu_\mu$
scattering even though such processes may have a larger rate than some
of the included leptonic processes.  Processes of this type do not
exchange energy between the neutrinos and the background medium.  They
are therefore not expected to affect the emerging fluxes and should
also have a minor effect on the emitted spectra.

\subsection{Importance of Different Processes}

\subsubsection{Accretion-Phase Model~I}
\label{sec:messer}

Our first goal is to assess the relative importance of different
energy-exchange processes for the $\nu_{\mu}$ transport.  As a first
example we begin with our Accretion-Phase Model~I. The results from
our numerical runs are summarized in
Table~\ref{tab:NumericalResultsMesser} where for each run we give
$\langle\epsilon\rangle_{\rm flux}$, our fit parameter $\alpha$
determined by Eq.~(\ref{eq:pinchingalpharelation}), and the pinching
parameter $p_{\rm flux}$ for the emerging flux spectrum, the
temperature and degeneracy parameter of an effective Fermi-Dirac
spectrum producing the same first two energy moments, and the
luminosity.

The first row contains the muon neutrino flux characteristics of the
original Boltzmann transport calculation by Messer.  To make a
connection to these results we ran our code with the same input
physics, i.e.\ $\nu_\mu\re^\pm$ scattering (s) and $\re^+\re^-$
annihilation (p).  There remain small differences between the original
spectral characteristics and ours. These can be caused by differences
in the implementation of the neutrino processes, by the limited number
of energy and angular bins in the Boltzmann solver, the coarser
resolution of the radial grid in our Monte Carlo runs, and by our
simple blackbody lower boundary condition. We interpret the first two
rows of Table~\ref{tab:NumericalResultsMesser} as agreeing
sufficiently well with each other that a detailed understanding of the
differences is not warranted. Henceforth we will only discuss
differential effects within our own implementation.

\begin{deluxetable}{lllllrrcrcrc}
\tablecaption{\label{tab:NumericalResultsMesser}
Monte Carlo results for Accretion-Phase Model~I.}
\tablewidth{0pt}
\tablehead{\multicolumn{5}{l}{Energy exchange}
&$\langle\epsilon\rangle_{\rm flux}$&$\langle\epsilon^2\rangle_{\rm
flux}$&$\alpha$&$p_{\rm flux}$&$T$&$\eta$&$L_\nu$}
\startdata
\multicolumn{5}{l}{original run}
              &17.5& 388.& 2.7& 0.97&  5.2&  1.1& 14.4\\  
--&--&s &p &--&16.6& 362.& 2.2& 1.01&  5.3& $-$0.3& 15.8\\
b &--&s &p &--&16.3& 351.& 2.1& 1.02&  5.4& $-$2.2& 19.1\\
b &--&--&p &--&17.8& 419.& 2.1& 1.02&  5.9& $-$1.9& 20.1\\
b &r &--&p &--&15.1& 285.& 3.0& 0.96&  4.3&  1.6& 18.6\\  
b &r &s &p &--&14.2& 255.& 2.8& 0.98&  4.2&  1.1& 14.8\\  
b &r &s &p &n &14.4& 264.& 2.7& 0.97&  4.3&  1.2& 17.6\\  
b &--&s &p &n &16.6& 358.& 2.3& 1.00&  5.2&  0.2& 21.7\\  
--&--&s &p &n &16.9& 369.& 2.4& 0.99&  5.3&  0.4& 20.2\\  
b &r &s &--&--&14.0& 251.& 2.6& 0.99&  4.3&  0.6& 13.1\\  
b &r &s &--&n &14.4& 263.& 2.7& 0.97&  4.3&  1.2& 17.0\\  
--&r &s &p &--&14.5& 265.& 2.8& 0.97&  4.3&  1.2& 13.0\\  
--&r &s &p &n &14.7& 269.& 3.1& 0.96&  4.2&  1.7& 16.8\\  
b &r &sn&p &n &14.3& 260.& 2.7& 0.97&  4.3&  1.2& 17.9\\  
\enddata
\tablecomments{For energy exchange, ``b'' refers to bremsstrahlung,
``r'' to recoil, ``s'' to scattering on electrons and positrons, 
``p'' to $\re^+ \re^-$ annihilation, ``n'' to $\nue\bar\nue$ 
annihilation, and ``sn'' to scattering on both e$^\pm$ and $\nue$,
$\bar\nue$.  We give $\langle\epsilon\rangle_{\rm flux}$ and $T$ in
MeV, $\langle\epsilon^2\rangle_{\rm flux}$ in MeV$^2$, and $L_\nu$ in
$10^{51}~\rm erg~s^{-1}$.} 
\end{deluxetable}

In the next row (bsp) we include nucleon bremsstrahlung which has the
effect of increasing the luminosity by a sizable amount without
affecting much the spectral shape. This suggests that bremsstrahlung
is important as a source for $\nu_\mu\bar\nu_\mu$ pairs, but that the
spectrum is then shaped by the energy-exchange in scattering with
$\re^\pm$. In the next row we switch off $\re^\pm$ scattering (bp) so
that no energy is exchanged except by pair-producing processes.  The
spectral energy indeed increases significantly.  However, the biggest
energy-exchange effect in the scattering regime is nucleon recoil.  In
the next two rows we include recoil (brp) and then additionally
$\re^\pm$ scattering (brsp), both lowering the spectral energies and
also the luminosities.

The picture of all relevant processes is completed by adding
$\nue\bar\nue$ pair annihilation (brspn), which is similar to
$\re^+\re^-$ pair annihilation, but a factor of 2--3 more important
(Buras et al.\ 2002).  The luminosity is again increased, an effect
which is understood in terms of our blackbody picture for the number
and energy spheres.  In the lower panel of
Fig.~\ref{fig:rthermrealistic} we see that $R_{\rm therm}$ moves to
larger radii once ``n'' is switched on, the radiating surface of the
``blackbody'' increases and more pairs are emitted.  For both ``p''
and ``n'' $R_{\rm therm}$ is strongly energy dependent and therefore
it is impossible to define a sharp thermalization radius.

Switching off ``r'' again (bspn) shows that also with ``n'' included,
``r'' really dominates the mean energy and shaping of the spectrum.  

To study the relative importance of the different pair processes, we
switch off the leptonic ones (row ``brs'') and compare this to only
the leptonic processes (row ``rspn'').  In this stellar model both
types contribute significantly.  Comparing then ``brsp'' with ``brsn''
shows that among the leptonic processes ``n'' is clearly more
important than ``p''.

The last row ``brsnpn'' includes in addition to all other processes
scattering on $\nue$ and $\bar\nue$.  It was already shown by Buras
et~al.\ (2002) that this process is about half as important as
scattering on $e^\pm$ and its influence on the neutrino flux and
spectra is negligible.  We show this case for completeness but do not
include scattering on $\nue$ and $\bar\nue$ for any of our further
models.

\begin{figure}[ht]
\columnwidth=7.5cm
\plotone{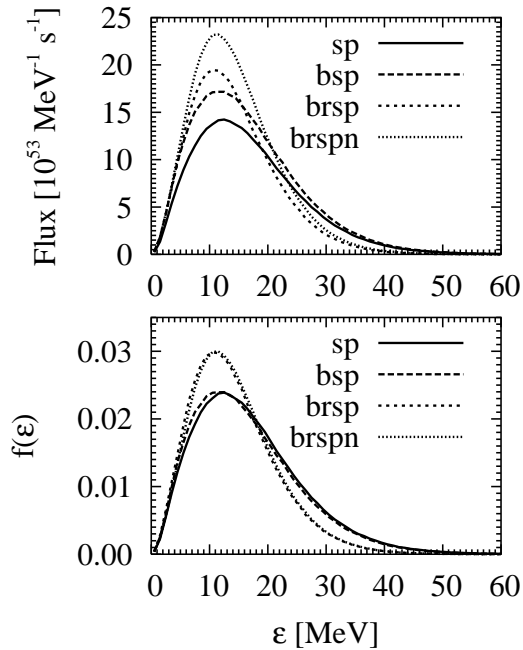}
\caption{\label{fig:ap1spectra} High-statistics spectra for
Accretion-Phase Model~I with different input physics as in
Table~\ref{tab:NumericalResultsMesser}.  {\it Upper
panel:}~Differential particle fluxes.  {\it Lower panel:}~Spectra
normalized to equal particle fluxes.}
\end{figure}

In order to illustrate some of the cases of
Table~\ref{tab:NumericalResultsMesser} we show in the upper panel of
Fig.~\ref{fig:ap1spectra} several flux spectra from high-statistics
Monte-Carlo runs.  Starting again with the input physics of the
original hydrodynamic simulation (sp) we add bremsstrahlung (b),
recoil (r), and finally $\nue\bar\nue$ pair annihilation (n).  Each of
these processes has a significant and clearly visible influence on the
curves. The pair-creation processes (``b'' and ``n'') hardly change
the spectral shape but increase the number flux, whereas recoil (r)
strongly modifies the spectral shape.  In the lower panel of
Fig.~\ref{fig:ap1spectra} we show the same curves, normalized to equal
particle fluxes.  In this representation it is particularly obvious
that the pair processes do not affect the spectral shape.

The very different impact of pair processes and nucleon recoils has a
simple explanation. The thermalization depth for the pair processes is
deeper than that of the energy-exchanging reactions, i.e.\ the
``number sphere'' is below the ``energy sphere.'' Therefore, the
particle flux is fixed more deeply in the star while the spectra are
still modified by energy-exchanging reactions in the scattering
atmosphere.

\subsubsection{Steep Power Law}

As another example we study the steep power-law model defined in
Eq.~(\ref{eq:powerlaws}) and Table~\ref{tab:powerlawmodels}.  This
model is supposed to represent the outer layers of a late-time
proto-neutron star but without being hydrostatically self-consistent.
It connects directly with Raffelt (2001), where the same profile was
used in a plane parallel setup, studying bremsstrahlung and nucleon
recoil.  The results of our runs are displayed in
Table~\ref{tab:NumericalResultsSteep} and agree very nicely with those
obtained by Raffelt (2001), corresponding to our cases ``b'' and
``br''.

\begin{deluxetable}{llllllrrcrcrc}
\tablecaption{\label{tab:NumericalResultsSteep}
Monte Carlo results for the steep power-law model.}
\tablewidth{0pt}
\tablehead{\multicolumn{5}{l}{Energy exchange}&$\Ye$
&$\langle\epsilon\rangle_{\rm flux}$&$\langle\epsilon^2\rangle_{\rm
flux}$&$\alpha$&$p_{\rm flux}$&$T$&$\eta$&$L_\nu$}
\startdata
        b &--&--&--&--&--- & 25.8& 962. & 1.2 & 1.11&---&---&  21.0\\     
        b &r &--&--&--&--- & 19.5& 487. & 2.6 & 0.98&6.0&0.7&  14.5\\ 
        b &--&--&p &--&0   & 25.4& 890. & 1.6 & 1.06&---&---&  23.8\\     
        b &--&--&p &--&0.05& 25.6& 908. & 1.6 & 1.06&---&---&  23.2\\     
        b &--&--&p &--&0.5 & 25.5& 917. & 1.4 & 1.08&---&---&  21.6\\     
        b &--&s &p &--&0   & 24.2& 787. & 1.9 & 1.03&---&---&  24.5\\     
        b &--&s &p &--&0.05& 23.8& 753. & 2.0 & 1.02&---&---&  24.5\\     
        b &--&s &p &--&0.5 & 21.3& 591. & 2.3 & 1.00&  6.8&$-$0.3&  23.1\\
        b &r &--&p &--&0.5 & 20.0& 507. & 2.7 & 0.98&  6.0&  1.0&  16.8\\ 
        b &r &s &p &--&0   & 20.3& 518. & 2.9 & 0.97&  5.9&  1.4&  19.7\\ 
        b &r &s &p &--&0.05& 20.3& 518. & 2.9 & 0.97&  5.9&  1.4&  19.5\\ 
        b &r &s &p &--&0.5 & 19.6& 488. & 2.7 & 0.98&  5.9&  1.1&  18.7\\ 
        b &r &s &p & n&0   & 20.7& 535. & 3.0 & 0.96&  5.8&  1.8&  23.9\\ 
b${\times}3$&r &s &p&n&0.05& 20.3& 522. & 2.7 & 0.97&  6.0&  1.3&  24.2\\ 
        b &r &s &p & n&0.05& 20.6& 530. & 3.0 & 0.96&  5.9&  1.7&  23.8\\ 
b${\times}0.3$&r&s&p&n&0.05& 20.7& 534. & 3.1 & 0.96&  5.8&  1.8&  23.4\\ 
        b &r &s &p & n&0.5 & 19.8& 499. & 2.7 & 0.97&  5.9&  1.2&  21.4\\ 
\enddata
\end{deluxetable}

\begin{deluxetable}{llllllrrcrcrc}
\tablecaption{\label{tab:NumericalResultsShallow}
Monte Carlo results for the shallow power-law model.}
\tablewidth{0pt}
\tablehead{\multicolumn{5}{l}{Energy exchange}&$\Ye$
&$\langle\epsilon\rangle_{\rm flux}$&$\langle\epsilon^2\rangle_{\rm
flux}$&$\alpha$&$p_{\rm flux}$&$T$&$\eta$&$L_\nu$}
\startdata
b &--&--&--&--&--- & 27.7& 1120.& 1.2 & 1.12&---&---&  20.3\\     
b &r &--&--&--&--- & 20.1&  521.& 2.5 & 0.99&  6.3&  0.4&  13.4\\ 
b &--&--&p &--&0   & 27.7&  974.& 2.7 & 0.98&  8.3&  1.0&  43.1\\ 
b &--&--&p &--&0.05& 27.9&  990.& 2.7 & 0.98&  8.3&  1.1&  43.3\\ 
b &--&--&p &--&0.5 & 28.3& 1019.& 2.7 & 0.98&  8.5&  1.0&  38.3\\ 
b &--&s &p &--&0   & 25.5&  830.& 2.6 & 0.98&  7.6&  1.1&  46.2\\ 
b &--&s &p &--&0.05& 25.4&  815.& 2.8 & 0.97&  7.5&  1.2&  46.3\\ 
b &--&s &p &--&0.5 & 23.5&  706.& 2.6 & 0.98&  7.1&  1.0&  44.8\\ 
b &r &--&p &--&0.5 & 22.5&  624.& 3.3 & 0.95&  6.1&  2.2&  33.1\\ 
b &r &s &p &--&0   & 22.3&  612.& 3.3 & 0.95&  6.1&  2.1&  39.6\\ 
b &r &s &p &--&0.05& 22.2&  609.& 3.2 & 0.95&  6.1&  2.1&  39.1\\ 
b &r &s &p &--&0.5 & 21.7&  585.& 3.1 & 0.95&  6.1&  1.9&  39.2\\ 
b &r &s &p & n&0   & 22.2&  608.& 3.3 & 0.94&  6.0&  2.2&  54.7\\ 
b &r &s &p & n&0.05& 22.4&  615.& 3.4 & 0.94&  6.1&  2.3&  54.9\\ 
b &r &s &p & n&0.5 & 21.8&  587.& 3.3 & 0.95&  6.1&  1.9&  51.3\\ 
\enddata
\end{deluxetable}

For investigating the importance of leptonic processes, we run our
code with a variety of neutrino interactions and in addition assume a
constant electron fraction $\Ye$ throughout the whole stellar 
atmosphere.  This
assumption is somewhat artificial, but gives us the opportunity to
study extreme cases in a controlled way.  In the relevant region $\Ye
= 0.5$ yields the highest possible electron density. In addition we
study the electron fraction being one order of magnitude smaller, $\Ye
= 0.05$, and finally the extreme case with an equal number of
electrons and positrons, $\Ye = 0$.

The first leptonic process we consider is $\re^+\re^-$ pair
annihilation.  Comparing the rows ``bp'' with the row ``b'' shows a
negligible effect on the spectrum, but a rise in luminosity.
Increasing $\Ye$ brings the luminosity almost back to the ``b'' case,
because the electron degeneracy rises and the positron density 
decreases so that the pair process becomes less important. 

Adding scattering on $\re^\pm$ forces the transported neutrinos to
stay closer to the medium temperature, i.e.\ reduces their mean energy.
Of course, the scattering rate increases with the number of electrons
and positrons, i.e.\ for higher $\Ye$ we get lower spectral energies.
For the luminosity the situation is more complicated.  Since the
neutrino flux energies decrease when we switch on $\re^\pm$ scattering
we would expect a lower luminosity.  However, the opacity of the
medium to neutrinos is strongly energy dependent and low energy
neutrinos can escape more easily than high-energy ones, increasing the
number flux.  On balance, the ``bsp'' luminosities are larger compared
to the ``bp'' ones.

To compare the scattering on $\re^\pm$ with that on nucleons, we
turn off ``s'' again and instead switch on recoil (r).  Qualitatively,
the energy exchange is very different from the earlier case.  In the
scattering on $\re^\pm$ a neutrino can exchange a large amount of
energy, while for scattering on nucleons the energy exchange is
small. But since neutrino-nucleon scattering is the dominant source of
opacity that keeps the neutrinos inside the star, the scatterings are
very frequent. This leads to a stronger suppression in the high-energy
tail of the neutrino spectrum and therefore to a visibly smaller mean
flux energy and lower effective spectral temperature, but higher
effective degeneracy. Many nucleon scatterings, however, are needed to
downgrade the high-energy neutrinos (different from e$^\pm$
scattering). Therefore neutrinos stay longer at high energies and
experience a larger opacity and a larger amount of
backscattering. This suppresses the neutrino flux significantly.
 
In the runs including both scattering reactions (brsp), we find a
mixture of the effects of e$^\pm$ and nucleon scatterings and an
enhanced reduction of the mean flux energy.

Finally, adding the neutrino pair process yields almost no change in
energy and pinching, but an increased luminosity as expected from the
analogous case in Sec.~\ref{sec:messer}.  Although this profile is
rather steep, leptonic pair processes are still important
(Fig.~\ref{fig:rthermsteep}).
 
In order to estimate the sensitivity to the exact treatment of nucleon
bremsstrahlung we have performed one run with the bremsstrahlung rate
artificially enhanced by a factor of 3, and one where it was decreased
by a factor 0.3. All other processes were included.  The emerging
fluxes and spectra indeed do not depend sensitively on the exact
strength of bremsstrahlung as argued in Sec.~\ref{sec:SimplePicture}.

\subsubsection{Shallow Power Law}

For the shallow power law almost the same discussion as for the steep
case applies.  As we can already infer from
Fig.~\ref{fig:rthermshallow}, leptonic processes are more important.
This leads to a much higher increase of the neutrino flux once ``p''
or ``n'' are included, and to stronger spectral pinching when
e$^+$e$^-$ annihilation is switched on. Scattering on $\re^\pm$
downgrades the transported neutrino flux by a larger amount.

\subsection{The Effect of Binning}

Evidently nucleon recoil plays an important role for the $\nu_\mu$
spectrum formation. It is straightforward to implement this reaction
in our Monte Carlo approach, but it may be more difficult in those
treatments of neutrino transport that rely on binned energy spectra.
The energy exchange in a given $\nu\rN$ collision is relatively small
(see e.g.~Raffelt 2001) so that one may lose this effect if the
spectrum is too crudely binned.  To test the impact of binning we have
performed two runs for the Accretion Phase Model~I where we fix the
neutrino energies to a small number of values. After every interaction
the final-state energy is set to the central value of the energy bin
it falls into.  We have chosen the 17 logarithmically spaced bins on
the interval from 0 to around 380~MeV that are used in the simulations
of the Garching SN group.

\begin{figure}[ht]
\columnwidth=7.5cm
\plotone{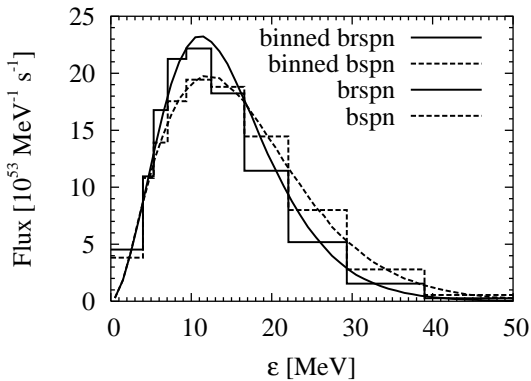}
\caption{\label{fig:bins} Comparison of runs with binned and unbinned
neutrino energies.}
\end{figure}

In Fig.~\ref{fig:bins} we compare the emergent flux spectra from runs
with binning (histograms) and without (smooth curves).  For both cases
recoil was once included (brspn) and once not (bspn).  The mean
energies calculated from either binned or unbinned runs agree to
better than 0.5\%, and also the shapes are well reproduced, although
there are slight differences around the spectral peak.  We conclude
that the impact of recoil is well accounted for in our runs with
discrete energies.  Therefore, the energy grid of the Garching group
should well suffice to reproduce the main spectral impact of nucleon
recoil.

\subsection{Detailed Spectral Shape}

\label{sec:High-Statistics}

Thus far we have characterized the neutrino spectra by a few simple
parameters.  However, it is extremely useful to have a simple analytic
fit to the overall spectrum that can be used, for example, to simulate
the response of a neutrino detector to a SN signal.  To study the
quality of different fit functions we have performed a few
high-statistics Monte-Carlo runs for the Accretion Phase Model~I,
including all interaction processes. Moreover, we have performed these
runs for the flavors $\nu_\re$, $\bar\nu_\re$, and $\nu_\mu$. (A
detailed discussion of the flavor dependence is deferred to Sec.~4.)

In order to get smooth spectral curves we have averaged the output of
70,000 time steps.  In addition we have refined the energy grid of the
neutrino interaction rates.  Both measures leave the previous results
unaffected but increase computing time and demand for memory
significantly.
  
\begin{figure}
\columnwidth=12cm
\plotone{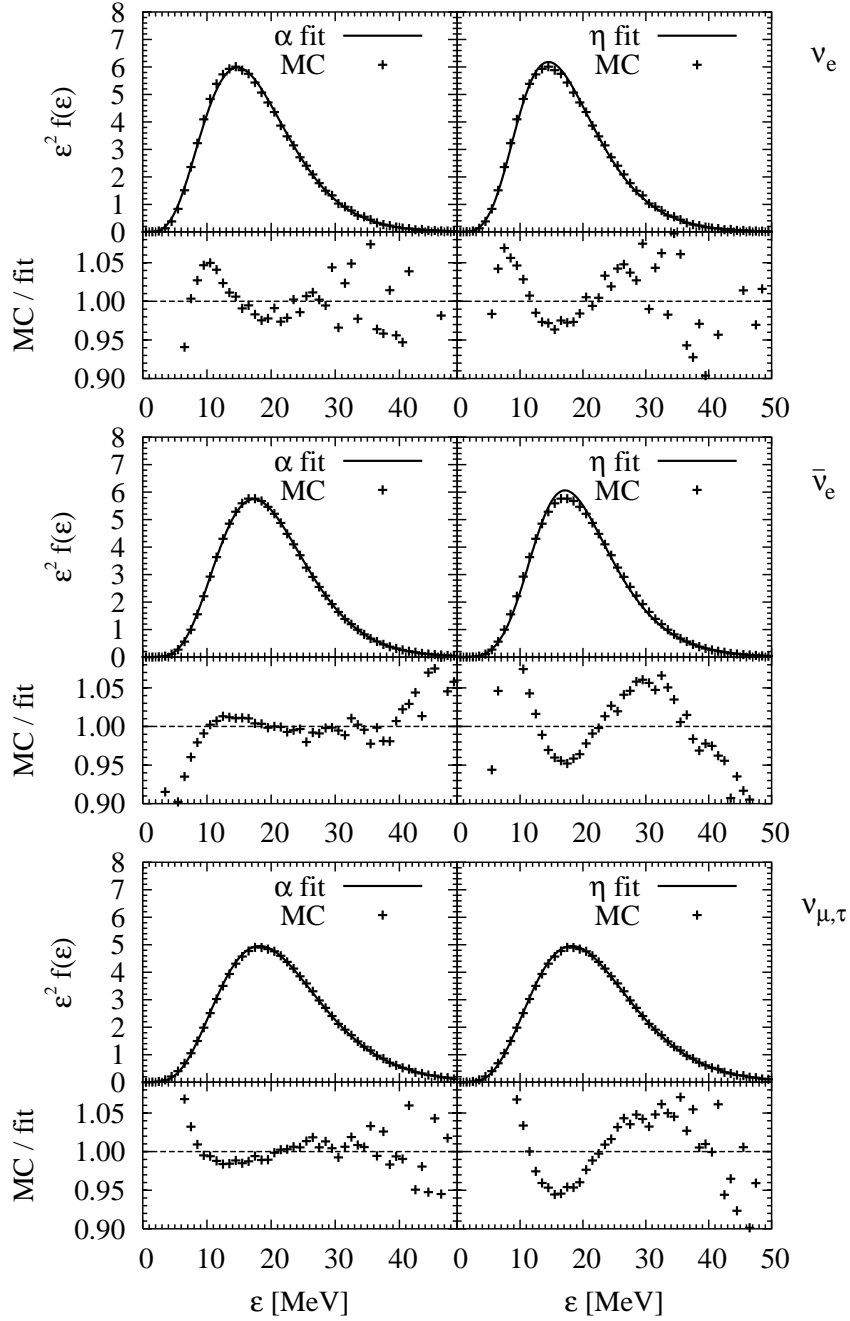}
\caption{\label{fig:highstatspecs} High-statistics spectra for
Accretion-Phase Model~I including all interaction processes.  The
Monte Carlo (MC) results are shown as crosses, the analytic fit
functions as smooth lines.  The left-hand panels use as fits
$f_\alpha(\epsilon)$ according to Eq.~(\ref{eq:powerlawfit}), the
right-hand panels $f_\eta(\epsilon)$ according to
Eq.~(\ref{eq:FermiDiracfit}).  Below the spectra we show the ratio
between Monte Carlo and fit.}
\end{figure}

In Fig.~\ref{fig:highstatspecs} we show our high-statistics Monte
Carlo (MC) spectra together with the $\alpha$-fit function
$f_\alpha(\epsilon)$ defined in Eq.~(\ref{eq:powerlawfit}) and the
$\eta$-fit function $f_\eta(\epsilon)$ of
Eq.~(\ref{eq:FermiDiracfit}).  The analytic functions can only fit the
spectrum well over a certain range of energies. We have chosen to
optimize the fit for the event spectrum in a detector, assuming the
cross section scales with $\epsilon^2$. Therefore, we actually show
the neutrino flux spectra multiplied with $\epsilon^2$.  Accordingly,
the parameters $\alpha$ and $\bar\epsilon$, as well as $\eta$ and $T$
and the normalizations are determined such that the energy moments
$\langle\epsilon^2\rangle$, $\langle\epsilon^3\rangle$, and
$\langle\epsilon^4\rangle$ are reproduced by the fits.

Below each spectrum we show the ratio of our MC results with the fit
functions.  In the energy range where the statistics in a detector
would be reasonable for a galactic SN, say from 5--10~MeV up to around
40~MeV, both types of fits represent the MC results nicely.  However,
in all cases the $\alpha$-fit works somewhat better than the
$\eta$-fit.

We have repeated this exercise for the steep power-law models with
$q=2.5$ and the one with $q=3.5$. The quality of the fits is
comparable to the previous example.

\subsection{Summary}

We find that the $\nu_\mu$ spectra are reasonably well described by
the simple picture of a blackbody sphere determined by the
thermalization depth of the nucleonic bremsstrahlung process, the
``filter effect'' of the scattering atmosphere, and energy transfers
by nucleon recoils. This is also true for the $\nu_\mu$ flux in case
of steep neutron-star atmospheres. For more shallow atmospheres pair
annihilation (e$^+$e$^-$ and $\nu_{\rm e}\bar\nu_{\rm e}$), however,
yields a large contribution to the emitted $\nu_{\mu}$ flux and
e$^\pm$ scattering reduces the mean flux energy significantly. It is
therefore important for state-of-the art transport calculations to
include these leptonic processes. The traditional process
$\re^+\re^-\to\nu_\mu\bar\nu_\mu$ is subdominant compared to
$\nue\bar\nue\to\nu_\mu\bar\nu_\mu$ as previously found
by Buras et~al.\ (2002).  The relative importance of the various
reactions depends on the stellar profile.

Neutrinos emitted from a blackbody surface and filtered by a
scattering atmosphere without recoils and leptonic processes have an
anti-pinched spectrum (Raffelt 2001). However, after all
energy-exchanging reactions have been included we find that the
spectra are always pinched. When described by effective Fermi-Dirac
distributions, the nominal degeneracy parameter $\eta$ is typically in
the range 1--2, depending on the profile and electron concentration.

\onecolumn
\twocolumn


\begin{deluxetable}{lllrccrcrr}
\tablecaption{\label{tab:MultiFlavor}
Comparing Monte Carlo results for different flavors.}
\tablewidth{0pt}
\tablehead{Model and Flavor
&$\Ye$&$\langle\epsilon\rangle_{\rm
flux}$&$\langle\epsilon^2\rangle_{\rm flux}$&$\displaystyle
\frac{\langle\epsilon\rangle_{\rm
flux}}{\langle\epsilon_{\bar\nue}\rangle_{\rm flux}}$&  
$\alpha$&$p_{\rm flux}$
&$T$&$\eta$&$L_\nu$}
\startdata
\multicolumn{9}{l}{\bf Accretion-Phase Model~I}\\
\multicolumn{9}{l}{\quad Original}\\
\qquad$\nu_\mu$, $\bar\nu_\mu$ 
                 &---&  17.5 &  388. & 1.20 &  2.7& 0.97&  5.2&  1.1&  14.4 \\ 
\qquad$\bar\nue$ &---&  14.6 &  253. & 1    &  4.4& 0.91&  3.5&  3.4&  29.2 \\ 
\qquad$\nue$     &---&  12.5 &  190. & 0.86 &  3.6& 0.93&  3.2&  2.8&  30.8 \\ 
\multicolumn{9}{l}{\quad Our runs}\\                                    
\qquad$\nu_\mu$, $\bar\nu_\mu$ (``sp'')                                 
                 &---&  16.6 &  362. & 1.19 &  2.2& 1.01&  5.3& $-$0.3&15.8 \\ 
\qquad$\nu_\mu$, $\bar\nu_\mu$ (``brspn'')                              
                 &---&  14.3 &  260. & 1.02 &  2.7& 0.97&  4.3&  1.2&  17.9 \\ 
\qquad$\nu_\mu$ (weak magnetism)                                   
                 &---&  14.2 &  254. & 1.01 &  2.9& 0.96&  4.1&  1.6&  17.4 \\ 
\qquad$\bar\nu_\mu$ (weak magnetism)                               
                 &---&  14.9 &  281. & 1.06 &  2.8& 0.97&  4.3&  1.5&  18.3 \\ 
\qquad$\bar\nue$ &---&  14.0 &  237. & 1    &  3.8& 0.93&  3.6&  2.7&  31.7 \\ 
\qquad$\nue$     &---&  11.8 &  175. & 0.84 &  2.9& 0.97&  3.4&  1.4&  31.9 \\ 
\multicolumn{9}{l}{\bf Accretion-Phase Model~II}\\                      
\multicolumn{9}{l}{\quad Original}\\                                    
\qquad$\nu_\mu$, $\bar\nu_\mu$                                          
                 &---&  17.2 &  380. & 1.09 &  2.5& 0.98&  5.2&  0.8&  32.4 \\ 
\qquad$\bar\nue$ &---&  15.8 &  300. & 1    &  4.0& 0.92&  4.0&  3.0&  68.1 \\ 
\qquad$\nue$     &---&  12.9 &  207. & 0.82 &  3.1& 0.96&  3.7&  1.7&  65.6 \\ 
\multicolumn{9}{l}{\quad Our runs}\\                                    
\qquad$\nu_\mu$, $\bar\nu_\mu$                                          
                 &---&  15.7 &  317. & 1.02 &  2.5& 0.98&  4.8&  0.8&  27.8 \\ 
\qquad$\bar\nue$ &---&  15.4 &  283. & 1    &  4.2& 0.92&  3.8&  3.2&  73.5 \\ 
\qquad$\nue$     &---&  13.0 &  207. & 0.84 &  3.4& 0.95&  3.6&  2.1&  73.9 \\ 
\multicolumn{9}{l}{\bf Steep Power Law $\mathbold{p=10}$}\\                               
\quad $q=2.5$\\                                                         
\qquad$\nu_\mu$, $\bar\nu_\mu$                                          
                &0.15 & 20.4 &  525. & 1.10 &  2.8& 0.96&  5.9&  1.5&  23.5 \\ 
\qquad$\bar\nue$&0.15 & 18.5 &  413. & 1    &  3.8& 0.92&  4.6&  3.0&  23.5 \\ 
\qquad$\nue$    &0.15 & 12.7 &  198. & 0.69 &  3.4& 0.94&  3.4&  2.4&  12.8 \\ 
\qquad$\nu_\mu$, $\bar\nu_\mu$                                          
                &0.2  & 20.4 &  521. & 1.14 &  3.0& 0.97&  5.9&  1.5&  23.3 \\ 
\qquad$\bar\nue$&0.2  & 17.9 &  383. & 1    &  4.1& 0.92&  4.4&  3.1&  11.7 \\ 
\qquad$\nue$    &0.2  & 13.4 &  218. & 0.75 &  3.7& 0.93&  3.4&  2.9&  24.4 \\ 
\quad$q=3.0$\\                                                          
\qquad$\nu_\mu$, $\bar\nu_\mu$                                          
                &0.1  & 17.7 &  393. & 1.14 &  2.9& 0.96&  5.0&  1.8&  12.7 \\ 
\qquad$\bar\nue$&0.1  & 15.5 &  289. & 1    &  3.9& 0.93&  4.0&  2.8&   8.8 \\ 
\qquad$\nue$    &0.1  & 10.5 &  132. & 0.68 &  4.1& 0.92&  2.6&  3.0&   6.6 \\ 
\quad$q=3.5$\\                                                          
\qquad$\nu_\mu$, $\bar\nu_\mu$                                          
                &0.07 & 15.8 &  310. & 1.22 &  3.1& 0.95&  4.4&  2.1&   7.9 \\ 
\qquad$\nu_\mu$ (weak magnetism)                                   
                &0.07 & 15.5 &  296. & 1.19 &  3.3& 0.94&  4.2&  2.3&   7.7 \\ 
\qquad$\bar\nu_\mu$ (weak magnetism)                               
                &0.07 & 16.5 &  337. & 1.27 &  3.2& 0.95&  4.5&  2.1&   8.3 \\ 
\qquad$\bar\nue$&0.07 & 13.0 &  207. & 1    &  3.4& 0.94&  3.5&  2.3&   4.3 \\ 
\qquad$\nue$    &0.07 &  9.4 &  103. & 0.72 &  5.0& 0.90&  2.1&  3.9&   4.1 \\ 
\multicolumn{9}{l}{\bf Shallow Power Law $\mathbold{p=5}$, $\mathbold{q=1}$}\\            
\qquad$\nu_\mu$, $\bar\nu_\mu$                                          
                &0.3  & 22.0 &  596. & 1.14 &  3.3& 0.94&  6.0&  2.2&  53.9 \\ 
\qquad$\bar\nue$&0.3  & 19.3 &  440. & 1    &  4.5& 0.91&  4.5&  3.7&  85.7 \\ 
\qquad$\nue    $&0.3  & 14.7 &  262. & 0.76 &  3.7& 0.93&  3.8&  2.7&  56.5 \\ 
\enddata
\tablecomments{We give $\langle\epsilon\rangle_{\rm flux}$ and $T$ in
MeV, $\langle\epsilon^2\rangle_{\rm flux}$ in MeV$^2$, $L_\nu$ in
$10^{51}~\rm erg~s^{-1}$.}  
\end{deluxetable}

\clearpage

\section{\uppercase{Comparing Different\\ Flavors}}

\subsection{Monte Carlo Study}
\label{sec:ourflavorcomp}

The new energy-exchange channels studied in the previous section lower
the average $\nu_{\mu}$ energies.  In order to compare the $\nu_\mu$
fluxes and spectra with those of $\nue$ and $\bar\nue$ we perform a
new series of runs where we include the full set of relevant
microphysics for $\nu_\mu$ and also simulate the transport of $\nue$
and $\bar\nue$.

The microphysics for the interactions of $\nue$ and $\bar\nue$ is the
same as in Janka \& Hillebrandt (1989a,b), i.e.\ charged-current
reactions of e$^\pm$ with nucleons, iso-energetic scattering on
nucleons, scattering on $\re^\pm$, and $\re^+\re^-$ pair annihilation.
In principle one should also include nucleon bremsstrahlung and the
effect of nucleon recoils for the transport of $\nue$ and $\bar\nue$,
but their effects will be minimal.  Therefore, we preferred to leave
the original working code unmodified for these flavors.

In the first three rows of Table~\ref{tab:MultiFlavor} we give the
spectral characteristics for the Accretion-Phase Model~I from the
original simulation of Messer.  The usual hierarchy of
average neutrino energies is found, i.e.\
$\langle\epsilon_{\nue}\rangle:\langle\epsilon_{\bar\nue}\rangle
:\langle\epsilon_{\nu_\mu}\rangle=0.86 : 1 : 1.20$. The luminosities
are essentially equal between $\bar\nue$ and $\nue$ while $\nu_\mu$,
$\bar\nu_\mu$, $\nu_\tau$, and $\bar\nu_\tau$ each provide about half
of the $\bar\nue$ luminosity.

Our Monte Carlo runs of this profile establish the same picture for
the same input physics. Although our mean energies are slightly offset
to lower values for all flavors relative to the original run, our
energies relative to each other are
$\langle\epsilon_{\nue}\rangle:\langle\epsilon_{\bar\nue}\rangle
:\langle\epsilon_{\nu_\mu}\rangle=0.84:1:1.19$ and thus very
similar. However, once we include all energy exchanging processes we
find $0.84:1:1.02$ instead.  Therefore,
$\langle\epsilon_{\nu_\mu}\rangle$ no longer exceeds
$\langle\epsilon_{\bar\nue}\rangle$ by much.  The luminosity of
$\nu_\mu$ is about half that of $\nue$ or $\bar\nue$ which are
approximately equal, in rough agreement with the original results.
Even though the additional processes lower the mean energy of
$\nu_\mu$ they yield a more than 10\% higher $\nu_{\mu}$ luminosity, 
mainly due to $\nue\bar\nue$ annihilation.

As another example of an accreting proto-neutron star we use the
Accretion-Phase Model~II.  The neutrino interactions included in this
model were nucleon bremsstrahlung, scattering on $\re^\pm$, and $\rm
e^+ \rm e^-$ annihilation.  Nucleon correlations, effective mass, and
recoil were taken into account, following Burrows \& Sawyer (1998,
1999), as well as weak magnetism effects (Horowitz 2002) and quenching
of $g_A$ at high densities (Carter \& Prakash 2002).  All these
improvements to the traditional microphysics affect mainly $\nu_{\mu}$
and to some degree also $\bar\nu_{\rm e}$.  Weak magnetism terms
decrease the nucleon scattering cross sections for $\bar\nu_{\mu}$
more strongly than they modify $\nu_{\mu}$ scatterings.  In this
hydrodynamic calculation, however, $\nu_{\mu}$ and $\bar\nu_{\mu}$
were treated identically by using the average of the corresponding
reaction cross sections. The effects of weak magnetism on the
transport of $\nu_{\mu}$ and $\bar\nu_{\mu}$ are therefore not
included to very high accuracy. Note, moreover, that the original data
come from a general relativistic hydrodynamic simulation with the
solution of the Boltzmann equation for neutrino transport calculated
in the comoving frame of the stellar fluid. Therefore the neutrino
results are affected by gravitational redshift and, depending on where
they are measured, may also be blueshifted by Doppler effects due to
the accretion flow to the nascent neutron star.

Our Monte Carlo simulation in contrast was performed on a static
background without general relativistic corrections. It includes
bremsstrahlung, recoil, $\rm e^+ \rm e^-$ pair annihilation,
scattering on $\rm e^\pm$, and $\nu_{\rm e}\bar\nu_{\rm e}$
annihilation, i.e.\ our microphysics is similar but not identical with
that used in the original run.  As an outer radius we took 100~km; all
flux parameters are measured at this radius because farther out
Doppler effects of the original model would make it difficult to
compare the results.  Keeping in mind that we use very different
numerical approaches and somewhat different input physics, the
agreement in particular for $\nu_{\rm e}$ and $\bar\nu_{\rm e}$ is
remarkably good.  This agreement shows once more that our Monte Carlo
approach likely captures at least the differential effects of the new
microphysics in a satisfactory manner.

\begin{figure}[b]
\columnwidth=6.5cm
\plotone{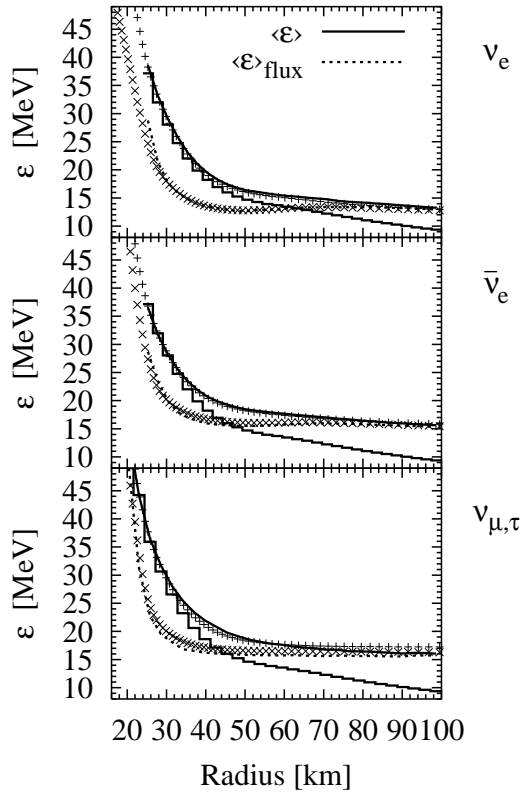}
\caption{\label{fig:ramppcomp} Comparison of the Accretion-Phase
Model~II calculations. Continuous lines show our Monte Carlo runs
while crosses represent the original simulation by Rampp. The steps
correspond to $\langle\epsilon\rangle = 3.15\,T$.}
\end{figure}

In Fig.~\ref{fig:ramppcomp} we compare our calculations for the
Accretion-Phase Model~II with those of the original simulation.  The
step-like curve again represents the mean energy of neutrinos in LTE
for zero chemical potential.  The smooth solid line is the mean energy
$\langle\epsilon\rangle$ from our runs, the dotted (lower) line gives
$\langle\epsilon\rangle_{\rm flux}$.  The crosses are the
corresponding results from the original runs.  For the transport of
$\nu_\mu$ our inner boundary is $R_{\rm in} = 16$~km, while for $\nue$
and $\bar\nue$ we use $R_{\rm in} = 24$~km.  For $\nue$ and $\bar\nue$
the charged-current processes (urca) keep these neutrinos in LTE up to
larger radii than pair processes in the case of $\nu_\mu$.  With our
choice of $R_{\rm in}$ the neutrinos are in LTE within the innermost
radial zones.

The results are similar to the Accretion-Phase Model~I.  The
luminosities are not equipartitioned but instead follow roughly
$L_{\nue} \approx L_{\bar\nue}\approx 2 \, L_{\nu_\mu}$.  The ratios
of mean energies are
$\langle\epsilon_{\nue}\rangle:\langle\epsilon_{\bar\nue}\rangle
:\langle\epsilon_{\nu_\mu}\rangle=0.82 : 1 : 1.09$ in the original run
and $0.84 : 1 : 1.02$ in our run.

In summary, both accretion-phase models agree reasonably well in the
$\langle\epsilon_{\nue}\rangle:\langle\epsilon_{\bar\nue}\rangle$
ratio for all runs. Moreover, using traditional input physics one
finds something like $\langle\epsilon_{\bar\nue}\rangle
:\langle\epsilon_{\nu_\mu}\rangle=1 : 1.20$. Depending on the
implementation of the new input physics and depending on the model one
finds results between $\langle\epsilon_{\bar\nue}\rangle
:\langle\epsilon_{\nu_\mu}\rangle=1 : 1.02$ and $1:1.09$.  The higher
ratio in Rampp's simulation could be due to the inclusion of weak
magnetism which tends to raise $\langle\epsilon_{\nu_\mu}\rangle$ more
than $\langle\epsilon_{\bar\nu_{\rm e}}\rangle$.

In order to estimate the corresponding results for later stages of the
proto-neutron star evolution we employ our steep power-law model.  We
vary the power $q$ of the temperature profile within a reasonable
range so that $q/p = 0.25$--0.35, with $q$ and $p$ defined in
Eq.~(\ref{eq:powerlaws}). $\Ye$ is fixed by demanding roughly equal
number fluxes for $\nue$ and $\bar\nue$ because a few seconds after
bounce deleptonization should be essentially complete.  The fluxes of
these neutrinos depend very sensitively on $\Ye$ so that this
constraint is only reached to within about 30\% without tuning $\Ye$
to three decimal places.  However, the mean energies are rather
insensitive to the exact value of $\Ye$.  This is illustrated by the
steep power-law model with $q = 2.5$ where we show results for
$\Ye=0.15$ and 0.20.  The number fluxes of $\nue$ and $\bar\nue$
differ by less than 30\% for $\Ye=0.15$, but differ by a factor of 3
for $\Ye=0.2$.  At the same time, the average spectral energies barely
change.

The ratios of mean energies are not very different from those of the
accretion-phase models.  Of course, the absolute flux energies have no
physical meaning because we adjusted the stellar profile in order to
obtain realistic values.  For the luminosities we find $L_{\nue}<
L_{\nu_\mu}$, different from the accretion phase. The steep power law
implies that the radiating surfaces are similar for all flavors so
that it is not surprising that the flavor with the largest energies
also produces the largest luminosity.

We find that $\langle\epsilon_{\nu_\mu}\rangle$ always exceeds
$\langle\epsilon_{\bar\nue}\rangle$ by a small amount, the exact value
depending on the stellar model.  During the accretion phase the
energies seem to be almost identical, later they may differ by up to
20\%. We have not found a model where the energies differ by the large
amounts which are sometimes assumed in the literature.  At late times
when $\Ye$ is small the microphysics governing $\bar\nue$ transport is
closer to that for $\nu_\mu$ than at early times.  Therefore, one
expects that at late times the behavior of $\bar\nue$ is more similar
to $\nu_\mu$ than at early times. We do not see any argument for
expecting an extreme hierarchy of energies at late times for
self-consistent stellar models.

We never find exact equipartition of the flavor-dependent
luminosities.  Depending on the stellar profile the fluxes can
mutually differ by up to a factor of 2 in either direction.

\subsection{Weak Magnetism}

Weak magnetism causes a significant correction to the neutrino-nucleon
cross section that arises due to the large anomalous magnetic moments
of protons and neutrons (Vogel \& Beacom 1999, Horowitz 2002).  It
increases the neutrino interaction rate but lowers the rate for
anti-neutrinos.  It is expected to be a small correction in the SN
context, but has never been implemented so far.  Following Horowitz
(2002) we add weak magnetism to our nucleon-recoil rate as given in
Appendix~\ref{appendix:rates}.

Our Monte Carlo code transports only one species of neutrinos at a
time. In order to test the impact of weak magnetism we assumed that a
chemical potential for $\nu_\mu$ would build up, and assumed a fixed
value for the $\nu_\mu$ degeneracy parameter throughout our stellar
model. We then iterated several runs for $\nu_\mu$ and $\bar\nu_\mu$
with different degeneracy parameters until their particle fluxes were
equal because in a stationary state there will be no net flux of
$\mu$-lepton number.

We performed this procedure for our Accretion-Phase Model I and our
steepest power law; the results are summarized in
Table~\ref{tab:MultiFlavor}.  In both cases the mean energies of
$\nu_\mu$ go down by $2\%$ and go up for $\bar\nu_\mu$ by $4\%$.  The
mean luminosities are unaffected. We conclude that weak-magnetism
corrections are small. Transporting $\nu_\mu$ and $\bar\nu_\mu$
separately in a self-consistent hydrodynamic simulation is probably
not worth the cost in computer time.

\subsection{Previous Literature}

\label{sec:PreviousLiterature}

There is a large recent body of literature quoted in our introduction
where the effect of flavor oscillations on SN neutrino spectra and
fluxes is studied. Many of these papers assumed that
$\langle\epsilon_{\nu_\mu}\rangle$ is much larger than
$\langle\epsilon_{\bar\nue}\rangle$ and that the luminosities between
all flavors were exactly equipartitioned. Our findings here are almost
orthogonal to this perception. Where does it come from?

To the best of our knowledge, the microphysics employed for $\nu_\mu$
transport is roughly the same in all published simulations. It
includes iso-energetic scattering on nucleons, $\re^+\re^-$
annihilation and $\nu_\mu\re^\pm$ scattering.  Of course, the
transport method and the numerical implementation of the neutrino
processes differ in the codes of different groups.  The new reactions
and nucleon recoil lower $\langle\epsilon_{\nu_\mu}\rangle$ and modify
the luminosities, but not by such a large amount as to explain a
completely different paradigm.  Therefore, we have inspected the
previous literature and collect a representative sample of pertinent
results in Table~\ref{tab:Literature}. Note that the simulations
discussed below did not in all cases use the same stellar models and
equations of state for the dense matter in the supernova core.

We begin with the simulations of the Livermore group who find robust
explosions by virtue of the neutron-finger convection
phenomenon. Neutrino transport is treated in the hydrodynamic models
with a multigroup flux-limited diffusion scheme.  Mayle, Wilson, \&
Schramm (1987) gave detailed results for their SN simulation of a
$25\,M_\odot$ star.  For half a second after bounce they obtained a
somewhat oscillatory behavior of the neutrino luminosities.  After the
prompt peak of the electron neutrino luminosity, they got
$L_{\nue}\approx L_{\bar\nue}\approx 2\, L_{\nu_\mu}\approx 50$--$130
\times 10^{51}~{\rm erg~s^{-1}}$.  After about one second the values
stabilize.  This calculation did not produce the ``standard''
hierarchy of energies. However, there is clearly a tendency that
$\bar\nue$ behave more similar to $\nu_\mu$ at late times.

\begin{deluxetable}{llrrrccrrrrr}
\tablecaption{\label{tab:Literature}
Flavor dependent flux characteristics from the literature.}
\tablewidth{0pt}
\tablehead{& tpb & $\langle\epsilon_{\nue}\rangle$ &
$\langle\epsilon_{\bar\nue}\rangle$ &
$\langle\epsilon_{\nu_\mu}\rangle$ & $\displaystyle
\frac{\langle\epsilon_\nu\rangle}{\langle\epsilon_{\bar\nue}\rangle}$ &
$L_{\nue}$ & $L_{\bar\nue}$ & $L_{\nu_\mu} $}
\startdata
Mayle et~al.\ (1987) &1.0 & 12 & 24 & 22 &  $0.50:1:0.92$ & 20
& 20 & 20\\ 
Totani et~al.\ (1998)&0.3 & 12 & 15 & 19 &  $0.80:1:1.26$ & 20
& 20 & 20\\ 
                     &10  & 11 & 20 & 25 &  $0.55:1:1.25$ & 0.5
& 0.5 & 1\\ 
Bruenn (1987)        &0.5 & 10 & 12 & 25 &  $0.83:1:2.08$ & 3
& 5 & 16\\ 
Myra \& Burrows (1990)&0.13&11 & 13 & 24 &  $0.85:1:1.85$ & 30
& 30 & 16\\ 
Janka \& Hillebrandt (1989b)
                     &0.3 &  8 & 14 & 16 &  $0.57:1:1.14$ & 30
& 220 & 65\\ 
Suzuki (1990)        & 1  & 9.5& 13 & 15 &  $0.73:1:1.15$ & 4
& 4 & 3\\ 
                     &20  & 8  & 10 &  9 &  $0.80:1:0.90$ & 0.3
& 0.3 & 0.07\\ 
Suzuki (1991)        & 1  & 9.5& 13 & 15 &  $0.73:1:1.15$ & 3
& 3 & 3\\ 
                     &15  & 8  &  9 & 9.5&  $0.89:1:1.06$ & 0.4
& 0.4 & 0.3\\ 
Suzuki (1993)        & 1  & 9  & 12 & 13 &  $0.75:1:1.08$ & 3
& 3 & 3\\ 
                     &15  & 7  &  8 &  8 &  $0.88:1:1.00$ & 0.3
& 0.3 & 0.3\\ 
Accretion-Phase Model~I (original) 
&0.32&13 & 15 & 18 &  $0.86:1:1.20$ & 31 & 29 & 14\\ 
Accretion-Phase Model~I (our run) 
&0.32&12 & 14 & 14 &  $0.84:1:1.02$ & 32 & 32 & 18\\ 
Accretion-Phase Model~II (original) 
&0.15&13 & 16 & 17 &  $0.82:1:1.09$ & 66 & 68 & 32\\ 
Accretion-Phase Model~II (our run) 
&0.15&13 & 15 & 16 &  $0.84:1:1.02$ & 74 & 74 & 28\\ 
Buras et al.\ (personal comm.) 
&0.25&14.1 & 16.5 & 16.8 &  $0.85:1:1.02$ & 43 & 44 & 32\\ 
\\
\multicolumn{9}{l}{\bf The following lines show
\boldmath{$\langle\epsilon\rangle_{\rm rms}$} instead of
\boldmath{$\langle\epsilon\rangle$}}\\ 
Mezzacappa et~al.\ (2001)&0.5 & 16 & 19 & 24 &  $0.84:1:1.26$ & 25
& 25 & 8\\ 
Liebend\"orfer et~al.\ (2001)&0.5 & 19 & 21 & 24 & $0.90:1:1.14$ & 30
& 30 & 10\\ 
\enddata
\tablecomments{We give the time post bounce (tpb) in s,
$\langle\epsilon\rangle$ in MeV, and $L_\nu$ in $10^{51}~\rm
erg~s^{-1}$.}
\end{deluxetable}

The most recent published Livermore simulation is a $20\,M_\odot$ star
(Totani et~al.\ 1998).  It shows an astonishing degree of luminosity
equipartition from the accretion phase throughout the early
Kelvin-Helmholtz cooling phase.  About two seconds after bounce the
$\nu_\mu$ flux falls off more slowly than the other flavors.  In
Table~\ref{tab:Literature} we show representative results for an early
and a late time. The mean energies and their ratios are consistent
with what we would have expected on the basis of our study.

With a different numerical code, Bruenn (1987) found for a
$25\,M_\odot$ progenitor qualitatively different results for
luminosities and energies.  At about 0.5~s after bounce the
luminosities and energies became stable at the values given in
Table~\ref{tab:Literature}.  This simulation is an example for an
extreme hierarchy of mean energies.

In Burrows (1988) all luminosities are said to be equal.  In addition
it is stated that for the first 5 seconds
$\langle\epsilon_{\nu_\mu}\rangle \approx 24$ MeV and the relation to
the other flavors is
$\langle\epsilon_{\nue}\rangle:\langle\epsilon_{\bar\nue}\rangle
:\langle\epsilon_{\nu_\mu}\rangle = 0.9:1:1.8$.  Detailed results are
only given for $\bar\nue$, so we are not able to add this reference to
our table.  The large variety of models investigated by Burrows (1988)
and the detailed results for $\bar\nue$ go beyond the scope of our
brief description. In a later paper Myra \& Burrows (1990) studied a
$13\,M_\odot$ progenitor model and found the extreme hierarchy of
energies shown in our table.

With the original version of our code Janka \& Hillebrandt (1989b)
performed their analyses for a $20\,M_\odot$ progenitor from a
core-collapse calculation by Hillebrandt (1987).  Of course, like our
present study, these were Monte Carlo simulations on a fixed
background model, not self-consistent simulations.  Taking into
account the different microphysics the mean energies are consistent
with our present work.  The mean energies of $\nu_{\rm e}$ were
somewhat on the low side relative to $\bar\nu_{\rm e}$ and the
$\bar\nu_{\rm e}$ luminosity was overestimated. Both can be understood
by the fact that the stellar background contained an overly large
abundance of neutrons, because the model resulted from a post-bounce
calculation which only included electron neutrino transport.

Suzuki (1990) studied models with initial temperature and density
profiles typical of proto-neutron stars at the beginning of the
Kelvin-Helmholtz cooling phase about half a second after bounce.  He
used the relatively stiff nuclear equation of state developed by
Hillebrandt \& Wolff (1985). In our table we show the results of the
model C12.  From Suzuki (1991) we took the model labeled C20 which
includes bremsstrahlung.  The model C48 from Suzuki (1993) includes
multiple-scattering suppression of bremsstrahlung. Suzuki's models are
the only ones from the previous literature which go beyond the
traditional microphysics for $\nu_\mu$ transport. It is reassuring
that his ratios of mean energies come closest to the ones we find.

Over the past few years, first results from Boltzmann solvers coupled
with hydrodynamic simulations have become available, for example the
unpublished ones that we used as our Accretion-Phase Models I and~II.
For convenience we include them in Table~\ref{tab:Literature}.
Further, we include a very recent accretion phase model of
the Garching group (Buras et al., personal communication) that
includes the full set of microphysical input.
Finally, we include two simulations similar to the Accretion-Phase
Model~I, one by Mezzacappa et~al.\ (2001) and the other by
Liebend\"orfer et~al.\ (2001).  These latter papers show rms energies
instead of mean energies. Recalling that the former tend to be about
45\% larger than the latter these results are entirely consistent with
our Accretion-Phase Models.  Moreover, the ratios of
$\langle\epsilon\rangle_{\rm rms}$ tend to exaggerate the spread
between the flavor-dependent mean energies because of different
amounts of spectral pinching, i.e.\ different effective degeneracy
parameters.  To illustrate this point we take the first two rows from
Table~\ref{tab:MultiFlavor} as an example.  The ratio of mean energies
for Fermi-Dirac spectra with temperatures $T_1=5.2$ and $T_2=3.5$ and
degeneracy parameters $\eta_1=1.1$ and $\eta_2=3.4$ is $\langle
\epsilon_1 \rangle / \langle \epsilon_2 \rangle = 1.19$, whereas the
ratio of rms energies equals~1.30.

To summarize, the frequently assumed exact equipartition of the
emitted energy among all flavors appears only in some simulations of
the Livermore group.  We note that the flavor-dependent luminosities
tend to be quite sensitive to the detailed atmospheric structure and
chemical composition.  On the other hand, the often-assumed extreme
hierarchy of mean energies was only found in the early simulations of
Bruenn (1987) and of Myra \& Burrows (1990), possibly a consequence of
the neutron-star equation of state used in these calculations.

If we ignore results which appear to be ``outliers'', the picture
emerging from Table~\ref{tab:Literature} is quite consistent with our
own findings.  For the luminosities, typically $L_{\nue} \approx
L_{\bar\nue}$ and a factor of \hbox{2--3} between this and
$L_{\nu_\mu}$ in either direction, depending on the evolutionary
phase.  For the mean energies we read typical ratios in the range of
$\langle\epsilon_{\nue}\rangle:\langle\epsilon_{\bar\nue}\rangle
:\langle\epsilon_{\nu_\mu}\rangle=0.8$--$0.9 : 1 : 1.1$--1.3.  The
more recent simulations involving a Boltzmann solvers show a
consistent behavior and will in future provide reliable information
about neutrino fluxes and spectra.


\section{\uppercase{Discussion and Summary}}

We have studied the formation of neutrino spectra and fluxes in a SN
core.  Using a Monte Carlo code for neutrino transport, we varied the
microscopic input physics as well as the underlying static
proto-neutron star atmosphere. We used two background models from
self-consistent hydrodynamic simulations, and several power-law
models with varying power-law indices for the density and temperature
and different values for the electron fraction $\Ye$, taken to be
constant.

The $\nu_\mu$ transport opacity is dominated by neutral-current
scattering on nucleons. In addition, there are number-changing
processes (nucleon bremsstrahlung, leptonic pair annihilation) and
energy-changing processes (nucleon recoil, $\nu_\mu\re^\pm$
scattering). The $\nu_\mu$ spectra and fluxes are roughly accounted
for if one includes one significant channel of pair production and one
for energy exchange in addition to $\nu_\mu{\rN}$ scattering. For
example, the traditional set of microphysics (iso-energetic
$\nu_\mu{\rN}$ scattering, $\re^+\re^-$ annihilation, and
$\nu_\mu\re^\pm$ scattering) yields comparable spectra and fluxes to a
calculation where pairs are produced by nucleon bremsstrahlung and
energy is exchanged by nucleon recoil. The overall result is quite
robust against the detailed choice of microphysics.

However, in state-of-the-art simulations where one aims at a precision
better than some 10--20\% for the fluxes and spectral energies, one
needs to include bremsstrahlung, leptonic pair annihilation,
neutrino-electron scattering, and energy transfer in neutrino-nucleon
collisions.  Interestingly, the traditional $\re^+\re^-$ annihilation
process is always much less important than $\nue\bar\nue$
annihilation, a point that we previously raised with our collaborators
(Buras et~al.\ 2002). None of the reactions studied here can be
neglected except perhaps the traditional $\re^+\re^-$ annihilation
process and $\nu_\mu\nue$ and $\nu_\mu\bar\nue$ scattering.  
 
The existing treatments of the nuclear-physics aspects of the
$\rN\rN\to\rN\rN\nu\bar\nu$ bremsstrahlung process are rather
schematic. We find, however, that the $\nu_\mu$ fluxes and spectra do
not depend sensitively on the exact strength of the bremsstrahlung
rate.  Therefore, while a more adequate treatment of bremsstrahlung
remains desirable, the final results are unlikely to be much affected.

The transport of $\nu_\mu$ and $\bar\nu_\mu$ is usually treated
identically.  However, weak-magnetism effects render the $\nu_\mu\rN$
and $\bar\nu_\mu\rN$ scattering cross sections somewhat different
(Horowitz 2002), causing a small $\nu_\mu$ chemical potential to build
up.  We find that the differences between the average energies of
$\nu_\mu$ and $\bar\nu_\mu$ are only a few percent and can thus be
neglected for most purposes.

Including all processes works in the direction of making the fluxes
and spectra of $\nu_\mu$ more similar to those of $\bar\nue$ compared
to a calculation with the traditional set of input physics.  During
the accretion phase the neutron-star atmosphere is relatively
expanded, i.e.\ the density and temperature gradients are relatively
shallow. Our investigation suggests that during this phase
$\langle\epsilon_{\nu_\mu}\rangle$ is only slightly larger than
$\langle\epsilon_{\bar\nue}\rangle$, perhaps by a few percent or 10\%
at most. This result agrees with the first hydrodynamic simulation
including all of the relevant microphysics except $\nue\bar\nue$
annihilation (Accretion-Phase Model II) provided to us by M.~Rampp.
For the luminosities of the different neutrino species one finds
$L_{\bar\nue}\sim L_{\nue}\sim 2\,L_{\nu_\mu}$.  The smallness of
$L_{\nu_\mu}$ is not surprising because the effective radiating
surface is much smaller than for~$\bar\nue$.

During the Kelvin-Helmholtz cooling phase the neutron-star atmosphere
will be more compact, the density and temperature gradients will be
steeper. Therefore, the radiating surfaces for all species will become
more similar. In this situation $L_{\nu_\mu}$ may well become larger
than $L_{\bar\nue}$. However, the relative luminosities depend
sensitively on the electron concentration. Therefore, without a
self-consistent hydrostatic late-time model it is difficult to claim
this luminosity cross-over with confidence.

The ratio of the spectral energies is most sensitive to the
temperature gradient relative to the density gradient. In our
power-law models we used $\rho\propto r^{-p}$ and $T\propto
r^{-q}$. Varying $q/p$ between 0.25 and 0.35 we find that
$\langle\epsilon_{\bar\nue}\rangle:\langle\epsilon_{\nu_\mu}\rangle$
varies between $1:1.10$ and $1:1.22$. Noting that the upper range for
$q/p$ seems unrealistically large we conclude that even at late times
the spectral differences should be small; 20\% sounds like a safe
upper limit. We are looking forward to this prediction being checked
in a full-scale self-consistent neutron-star evolution model with a
Boltzmann solver.

The statements in the previous literature fall into two classes. One
group of workers, using the traditional set of microphysics, found
spectral differences between $\bar\nue$ and $\nu_\mu$ on the 25\%
level, a range which largely agrees with our findings in view of the
different microphysics.  Other papers claim ratios as large as
$\langle\epsilon_{\bar\nue}\rangle:\langle\epsilon_{\nu_\mu}\rangle
=1:1.8$ or even exceeding $1:2$.  We have no explanation for these
latter results. At least within the framework of our simple power-law
models we do not understand which parameter could be reasonably
adjusted to reach such extreme spectral differences.

In a high-statistics neutrino observation of a future galactic SN one
may well be able to discover signatures for flavor oscillations.
However, when studying these questions one has to allow for the
possibility of very small spectral differences, and conversely, for
the possibility of large flux differences. This situation is almost
orthogonal to what often has been assumed in papers studying possible
oscillation signatures. A realistic assessment of the potential of a
future galactic SN to disentangle different neutrino mixing scenarios
should allow for the possibility of very small spectral differences
among the different flavors of anti-neutrinos. The spectral
differences between $\nu_e$ and $\nu_{\mu,\tau}$ are always much
larger, but a large SN neutrino (as opposed to anti-neutrino) detector
does not exist.

The diffuse neutrino flux from all past SNe in the universe is
difficult to detect, although Super-Kamiokande has recently
established an upper limit that touches the upper end of theoretical
predictions (Malek et al.\ 2002).  If our findings are correct,
neutrino oscillations will not much enhance the high-energy tail of
the spectrum and thus will not significantly enhance the event rate.


\section*{\uppercase{Acknowledgments}}

We thank the Institute for Nuclear Theory (University of Washington,
Seattle) for its hospitality during a visit when this work was begun. 
In Munich, this work was partly supported by the Deut\-sche
For\-schungs\-ge\-mein\-schaft under grant No.\ SFB 375 and by the ESF
network Neutrino Astrophysics. We thank Bronson Messer and Markus
Rampp for providing unpublished stellar profiles from self-consistent
collapse simulations.


\appendix

\section{\uppercase{Monte Carlo Code}}

\label{app:MonteCarlo}

Our Monte Carlo code is based on that developed by Janka (1987) where
a detailed description of the numerical aspects can be found.  The
code was first applied to calculations of neutrino transport in
supernovae by Janka \& Hillebrandt (1989a,b) and Janka (1991).  It
uses Monte Carlo methods to follow the individual destinies of sample
neutrinos (particle ``packages'' with suitably attributed weights to
represent a number of real neutrinos) on their way through the star
from the moment of creation or inflow to their absorption or escape
through the inner or outer boundaries. The considered stellar
background is assumed to be spherically symmetric and static, and the
sample neutrinos are characterized by their weight factors and by
continuous values of energy, radial position and direction of motion,
represented by the cosine of the angle relative to the radial
direction. The rates of neutrino interactions with particles of the
stellar medium can be evaluated by taking into account Fermion
blocking effects according to the local phase-space distributions of
neutrinos (Janka \& Hillebrandt 1989b).

As background stellar models we use the ones described in
Sec.~\ref{sec:NeutronStarAtmospheres}. They are defined by radial
profiles of the density $\rho$, temperature $T$, and electron fraction
$\Ye$, i.e.\ the number of electrons per baryon.  The calculations
span the range between some inner radius $R_{\rm in}$ and outer radius
$R_{\rm out}$. These bound the computational domain which is divided
into 30 equally spaced radial zones.  In each zone $\rho$, $T$, and
$\Ye$ are taken to be constant.  $R_{\rm in}$ is chosen at such high
density and temperature that the neutrinos are in LTE in at least the
first radial zone. $R_{\rm out}$ is placed in a region where the
neutrinos essentially stream freely.  At $R_{\rm in}$ neutrinos are
injected isotropically according to LTE.  While a small net flux
across the inner boundary develops, the neutrinos emerging from the
star are generated almost exclusively within our computational
domain. If $R_{\rm in}$ is chosen so deep that the neutrinos are in
LTE, the assumed boundary condition for the flux will therefore not
affect the results.

The stellar medium is assumed to be in thermodynamic equilibrium with
nuclei being completely disintegrated into free nucleons.  Based on
$\rho$, $T$, and $\Ye$ we calculate all the required thermodynamic
quantities, notably the number densities, chemical potentials, and
temperatures of protons, neutrons, electrons, positrons, and the
relevant neutrinos.  The chemical potentials for $\nu_\mu$ and
$\nu_\tau$ are taken to be zero.  Next we compute the interaction
rates in each radial zone for all included processes.  In the
simulations discussed in the present work, fermion phase-space
blocking is calculated from the neutrino equilibrium distributions
instead of the computed phase-space distributions.  This
simplification saves a lot of CPU time because otherwise the rates
have to be re-evaluated whenever the distribution of neutrinos has
changed after a transport time step.  The approximation is justified
because phase-space blocking is most important in regions where
neutrinos frequently interact and thus are close to LTE.

At the start of a Monte Carlo run, 800,000 test neutrinos are randomly
distributed in the model according to the local equilibrium
distributions.  Each test neutrino represents a certain number of real
neutrinos.  In this initial setup the number of real neutrinos is
determined by LTE.  Then transport is started.  The time step is fixed
at $\Delta t = 10^{-7} s$; recall that the interaction rates do not
change.  At the beginning of each step neutrino creation takes place.
The number of test particles that can be created is given by the
number of neutrinos that were lost through the inner and outer
boundaries plus those absorbed by the medium.  Based on $\Delta t$,
the production rates, and the fact that the inner boundary radiates
neutrinos, we calculate the number of neutrinos that are produced in
one time step and distribute them among the available test neutrinos
by attributing suitable weight factors. The sample particles are
created within the medium or injected at the inner boundary in
appropriate proportions.

During a time step the path of each test particle through the stellar
atmosphere is followed by Monte Carlo sampling.  With random numbers
we decide whether it flies freely or interacts.  If it interacts it
can scatter or it can be absorbed; in this case we turn to the next
particle.  For scattering we determine the new momentum and position
and continue with the process until the time step is used up.
Particles leaving through the lower or upper boundaries are eliminated
from the transport.

After a certain number of time steps (typically around 15,000) the
neutrino distribution reaches a stationary state and further changes
occur only due to statistical fluctuations. At that stage we start
averaging the output quantities over the next 500 time steps.


\section{\uppercase{Neutrino Processes}}

\label{appendix:rates}

\subsection{Neutrino-Nucleon Scattering}

The rates for the $\nu_\mu\rN$ reactions are calculated following
Raffelt (2001).  For a neutrino with initial energy $\epsilon_1$ and
final energy $\epsilon_2$, the differential cross section is given by
\be\label{eq:nucscat} \frac{d\sigma}{d\epsilon_2\,d\cos\theta}
=\frac{C_A^2(3-\cos\theta)}{2\pi}\,G_F^2\,\epsilon_2^2\,
\frac{S(\omega,k)}{2\pi} \ee with $\omega = \epsilon_1 - \epsilon_2$,
$k$ the modulus of the momentum transfer to the medium, and $\theta$
the scattering angle.  We do not distinguish between protons and
neutrons.  Since for nonrelativistic nucleons the scattering cross
section is proportional to $C_V^2+3C_A^2$, the vector current
($C_V=-\frac{1}{2}$ for neutrons and $\frac{1}{2}-2\sin^2\Theta_{\rm
W}$ for protons) is small compared to the axial component, where we
use $|C_A| = 1.26/2$.  Neglecting the vector part simplifies the
calculations significantly and certainly has a smaller effect on the
scattering rates than other uncertainties, for example the in-medium
value of the coupling constants themselves.

In all of our runs without recoil the structure function is given by
\be 
S_{\rm no-recoil}(\omega,k)=2\pi\delta(\omega) \; .
\ee
This corresponds to infinitely heavy nucleons and represents the
traditional approximation in all previous simulations.  For the more
realistic case of recoiling nucleons the structure function 
becomes
\be 
\label{eq:s_recoil}
S_{\rm recoil}(\omega,k)=\sqrt{\frac{\pi}{\omega_k T}} \exp \left (
-\frac{\omega - \omega_k}{4 T \omega_k} \right )
\ee
with $\omega_k = k^2/2m$.  

Multiplying Eq.~(\ref{eq:nucscat}) with the density of nucleons,
ignoring phase space blocking of the essentially nondegenerate
nucleons, yields the differential rates that can then be integrated
for obtaining the required energy and angular differential rates.  In
the case of recoil the numerical integrations are rather tricky
because Eq.~(\ref{eq:nucscat}) is strongly forward peaked.  In our
code we employ the ``rejection method'' for obtaining the integrated
rates (Press et al.\ 1992).

\subsection{Weak Magnetism}

In order to account for weak magnetism effects, we multiply a
correction factor to our recoil amplitude, motivated by Horowitz
(2002).  The correction factor has the form
\begin{equation}
\label{eq:weakmag_corr}
\left ( 1 \pm \frac{4 C_A (C_V + F_2)}{C_A^2(3-\cos\theta)} \frac{k}{2
m} \right )^2 \, ,
\end{equation} 
with $F_2 = 1.019$ for protons and $F_2 = -0.963$ for neutrons.  The
upper sign is for neutrinos and the lower sign for anti-neutrinos.
Expression (\ref{eq:weakmag_corr}) is always positive; to first order
in $k/m$ it is given by 
\begin{equation}
1 \pm \frac{4 C_A (C_V +
F_2)}{C_A^2(3-\cos\theta)} \frac{k}{m} \,, 
\end{equation} 
and thus corresponds to the weak magnetism correction of Horowitz
(2002), except for one additional simplification.  In order to keep
our prescription for the recoil-amplitude via the structure function
(\ref{eq:s_recoil}) our correction factor must not contain any energy
dependence except for the momentum transfer $k$.  Therefore, we have
substituted $\epsilon_1 \cos\theta$, taken in the rest frame of the
nucleon, by our momentum transfer $k$.  This is correct for forward
and backward scattering, but only an approximation for other angles.

\subsection{Bremsstrahlung}

We also follow Raffelt (2001).  The mfp for the absorption of a
$\nu_\mu$ by inverse bremsstrahlung
$\rN\rN\nu_\mu\bar\nu_\mu\to\rN\rN$ is given by 
\bea
\lambda^{-1}_{\rm brems}&=&
\frac{C_A^2 G_F^2}{2}\, n_B\, \frac{1}{2\epsilon}\nonumber\\
&\times& \int\frac{d^3\bar k}{2\bar\epsilon\,(2\pi)^3}\,
f(\bar\epsilon)\,24\,\epsilon\,\bar\epsilon\,
S(\epsilon{+}\bar\epsilon)\,.\nonumber\\ 
\eea 
The over-barred quantities belong to the $\bar\nu_\mu$ that is
absorbed together with the primary $\nu_\mu$.  The occupation numbers
are taken to follow an equilibrium distribution with zero chemical
potential, and $\abs{C_A} = 1.26/2$ as in the scattering case.

\subsection{Pair Annihilation}

We now turn to $\re^+ \re^- \to \nu_\mu \bar\nu_\mu$ and $\nue
\bar\nue\to \nu_\mu \bar\nu_\mu$.  The matrix elements for both
processes are identical up to coupling constants while the phase-space
integrations only differ by the chemical potentials.

\begin{figure}[ht]
\columnwidth=4.5cm
\plotone{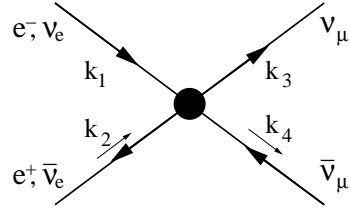}
\caption{\label{fig:feyn} Pair annihilation processes producing
$\nu_\mu\bar\nu_\mu$ pairs.}
\end{figure}

After summing over all spins and neglecting the rest masses, the squared
matrix element is
\bea 
\sum_{\rm spins}|{\cal M}|^2 &\!\!=\!\!&
8~G_F^2 \Bigl[(C_V+C_A)^2u^2 
\nonumber\\
&&\kern2em{}+(C_V-C_A)^2t^2\Bigr]
\label{eq:matrixel} 
\eea
with the Mandelstam variables $t=-2k_1\cdot k_3$ and $u=-2k_1\cdot
k_4$. The momenta are assigned to the particles as indicated in
Fig.~\ref{fig:feyn}.  The weak interaction constants for  
$\re^+\re^-$ annihilation are
\be 
\label{eq:weakcoefel}
C_V = -\frac{1}{2}+2\sin^2\Theta_{\rm W}\; ,\;\;\; C_A = -\frac{1}{2}
\ee
while for $\nue \bar\nue$ annihilation they are
\be 
\label{eq:weakcoefnu}
C_A = C_V = \frac{1}{2} \; .
\ee

For the interaction rates we have to perform the phase-space
integrations, using blocking factors for the final states and
occupation numbers for initial-state particles (Hannestad \& Madsen
1995, Yueh \& Buchler 1976).  Three integrations remain that can not
be carried out analytically.

Mu- and tau-leptons are almost absent in proto-neutron star
atmospheres so that the chemical potentials of the corresponding
neutrinos can be set to zero.  For the $\re^+\re^-$
reactions the local value of $\mu_{\re^-}=-\mu_{\re^+} \equiv \mu_\re$
can be obtained from $\rho$, $T$, and $\Ye$ by inverting
\be 
\frac{n_{\re^-}(\mu_\re)-n_{\re^+}(\mu_\re)}{n_{\rm baryons}} 
= \Ye  \; ,
\ee
where $n_{\re^-}(\mu_\re)$ and $n_{\re^+}(\mu_\re)$ are Fermi
integrals.  For $\nue$ and $\bar\nue$ the chemical potential is
obtained by the relation
\be 
\mu_{\nue} = - \mu_{\bar\nue} = \mu_\re + \mu_\rp - \mu_\rn
\ee
with the chemical potentials $\mu_\rp$ and $\mu_\rn$ of protons and
neutrons, respectively.

For $\nue \bar\nue$ annihilation we make use of the fact that the 
energy sphere of
$\nu_{\mu}$ lies always deeper inside the star than the $\nue$ and
$\bar\nue$ spheres (see Fig.~\ref{fig:rthermrealistic}).  Thus $\nue$
and $\bar\nue$ are in LTE and are part of the medium as far as the
transport of $\nu_\mu$ is concerned.  This approximation breaks down
at larger radii where this process is unimportant anyway.

For our numerical implementation we normally use Fermi-Dirac
statistics, but in order to reduce computation time one of the
remaining three phase-space integrations is approximated by the
analytic expressions given in Takahashi, El~Eid, \& Hillebrandt (1978).
This also requires simplifying the blocking factors.  With $\mu_\re =
-\mu_{\re^+} \ge 0$ we can approximate the positron occupation number
by a Maxwell-Boltzmann distribution.  For $\mu_\re/T \gsim 2$ this
holds to very good accuracy.  The greatest deviation is at $\mu_\re/T
=0$ and yields blocking factors too low by about 10\%.  However,
$\re^-$ and $\nue$ are always degenerate in the relevant regions.

\subsection{Scattering on Electrons and Electron Neutrinos}

\begin{figure}[ht]
\columnwidth=4.5cm
\plotone{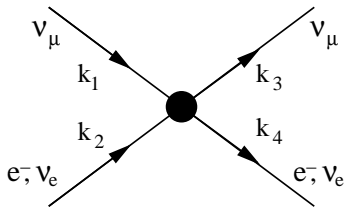}
\caption{\label{fig:feynsc} Leptonic scattering processes.}
\end{figure}

The matrix elements for these reactions are just the crossed
versions of the leptonic pair processes, 
\bea 
\label{eq:matrixelsc}
\sum_{\rm spins}|{\cal M}|^2 &\!\!=\!\!&
8~G_F^2 \Bigl[(C_V+C_A)^2s^2 
\nonumber\\
&&\kern1.5em{}+(C_V-C_A)^2u^2\Bigr]
\eea
with the same weak interaction coefficients of 
Eqs.~(\ref{eq:weakcoefel}) or
(\ref{eq:weakcoefnu}) for scattering on $\re^-$ or on $\nue$,
respectively.  For $s=2k_1\cdot k_2$ and $u=-2k_1\cdot k_4$ the
momenta are assigned to the particles according to
Fig.~\ref{fig:feynsc}.  Crossing the matrix element
Eq.~(\ref{eq:matrixelsc}) again by interchanging $u \lra t$, we obtain
scattering on $\re^+$ or $\bar\nue$. This is also true for scattering
of $\bar \nu_\mu$ on $\re^-$ or $\nue$; scattering of $\bar\nu_\mu$ on
$\re^+$ or $\bar\nue$ brings us back to Eq.~(\ref{eq:matrixelsc}).
For calculating the rates in our code we apply the same approximations
as in the previous section.

\end{document}